\documentclass[lettersize,journal]{IEEEtran}
\usepackage{amsmath,amsfonts}
\usepackage{algorithmic}
\usepackage{algorithm}
\usepackage{array}
\usepackage{textcomp}
\usepackage{stfloats}
\usepackage{url}
\usepackage{verbatim}
\usepackage{graphicx}
\usepackage{amssymb} 
\usepackage{amsfonts}
\usepackage{amsmath} 
\usepackage{physics} 
\usepackage{algorithmic}  
\usepackage{algorithm} 
\usepackage{graphicx}
\usepackage{color} 
\usepackage{bm}
\usepackage{epstopdf} 
\usepackage{subfigure}
\usepackage{booktabs}  
\usepackage[dvipsnames]{xcolor} 
\usepackage{cite} 
\usepackage{balance}
\usepackage{setspace}  
\usepackage{amsthm}
\usepackage{amsmath} 
\usepackage{hyperref}
\usepackage{balance}
\usepackage{amssymb}
\usepackage{placeins}
\theoremstyle{note} 

\newtheorem{remark}{Remark} 

\newcommand{\algref}[1]{\textbf{Algorithm \ref{#1}}}

\newcommand{\figref}[1]{Fig. \ref{#1}}
\newcommand{\secref}[1]{Section \ref{#1}}

\begin{document}
	
	\title{Extract the Best, Discard the Rest: \\CSI Feedback with Offline Large AI Models}
	
	\author{Jialin Zhuang, \textit{Student Member}, \textit{IEEE}, Yafei Wang, \textit{Graduate Student Member}, \textit{IEEE}, \\Hongwei Hou, \textit{Graduate Student Member}, \textit{IEEE}, Yu Han, \textit{Member}, \textit{IEEE}, Wenjin Wang, \textit{Member}, \textit{IEEE}, \\Shi Jin, \textit{Fellow}, \textit{IEEE}, Jiangzhou Wang, \textit{Fellow}, \textit{IEEE}
		\thanks{Jialin Zhuang, Yu Han, Shi Jin are with the National Mobile Communications Research Laboratory, Southeast University, Nanjing 210096, China (e-mail: zhuangjl0221@seu.edu.cn; hanyu@seu.edu.cn; jinshi@seu.edu.cn).}
		\thanks{Yafei Wang, Hongwei Hou, Wenjin Wang, and Jiangzhou Wang are with the National Mobile Communications Research Laboratory, Southeast University, Nanjing 210096, China, and also with Purple Mountain Laboratories, Nanjing 211100, China (e-mail: wangyf@seu.edu.cn; hongweihou@seu.edu.cn; wangwj@seu.edu.cn; j.z.wang@seu.edu.cn).}
		}
	
	\markboth{}%
	{Shell \MakeLowercase{\textit{et al.}}: A Sample Article Using IEEEtran.cls for IEEE Journals}
	
	
	\maketitle
	
	\bstctlcite{IEEEexample:BSTcontrol}
	\begin{abstract}
		Large AI models (LAMs) have shown strong potential in wireless communication tasks, but their practical deployment remains hindered by latency and computational constraints.
		In this work, we focus on the challenge of integrating LAMs into channel state information (CSI) feedback for frequency-division duplex (FDD) massive multiple-intput multiple-output (MIMO) systems. To this end, we propose two offline frameworks, namely site-specific LAM-enhanced CSI feedback (SSLCF) and multi-scenario LAM-enhanced CSI feedback (MSLCF), that incorporate LAMs into the codebook-based CSI feedback paradigm without requiring real-time inference. 
        Specifically, SSLCF generates a site-specific enhanced codebook through fine-tuning on locally collected CSI data, while MSLCF improves generalization by pre-generating a set of environment-aware codebooks. 
        Both of these frameworks build upon the LAM with vision-based backbone, which is pre-trained on large-scale image datasets and fine-tuned with CSI data to generate customized codebooks.
        This resulting network named LVM4CF captures the structural similarity between CSI and image, allowing the LAM to refine codewords tailored to the specific environments.
        To optimize the codebook refinement capability of LVM4CF under both single- and dual-side deployment modes, we further propose corresponding training and inference algorithms.
		Simulation results show that our frameworks significantly outperform existing schemes in both reconstruction accuracy and system throughput, without introducing additional inference latency or computational overhead. These results also support the core design methodology of our proposed frameworks, extracting the best and discarding the rest, as a promising pathway for integrating LAMs into future wireless systems.
	\end{abstract}
	

	%
	\IEEEpeerreviewmaketitle

	\section{Introduction}
	\IEEEPARstart{D}{riven} by the increasing demand for high data rates and spectral efficiency, massive multiple-input multiple-output (MIMO) technology has consistently been recognized as a key enabler in wireless network development \cite{MIMO_overview, MIMO_overview2}. However, fully unlocking the potential of massive MIMO systems requires accurate channel state information (CSI) at the base station (BS) \cite{CSI_importance, TENN, up_estimation}. In frequency-division duplex (FDD) systems, the absence of uplink-downlink channel reciprocity presents a major challenge, as user equipments (UEs) must estimate and feed back the downlink CSI to the BS. This feedback overhead scales with the size of the BS antenna array and becomes a critical bottleneck as the array aperture expands in future networks.
	
	To tackle this bottleneck, conventional model-driven approaches—such as codebook-based and compressed sensing (CS)-based feedback—have been explored as pioneering efforts to reduce CSI feedback overhead. Specifically, codebook-based feedback, standardized by 3rd Generation Partnership Project (3GPP) \cite{codebook_based,codebook_based2,codebook_based3}, allows UEs to quantize CSI by selecting codewords from a predefined codebook. Despite its widespread adoption, it suffers from the fundamental trade-off between feedback overhead and quantization errors, limiting its scalability in next-generation communication networks \cite{codebook_based3}. To alleviate this trade-off, CS-based feedback leverages the inherent sparsity \cite{sparsity} of channels to compress the channel into a low-dimensional representation \cite{CS_based_1, CS_based_2, CS3, CS4}, but always suffers from performance degradation in practicale propagation environments due to strong sparsity assumptions \cite{CS_based_3}.
	
	In response, data-driven approaches based on deep learning (DL) have emerged as promising alternatives, since they can bypass traditional model assumptions by learning compact and efficient CSI representations directly from data \cite{csinet, CsiNet_ext1, starris,autocsinet, CsiNet+, Attention-CSI, TransNet1, TransNet2,FBnet,DualNet,CovNet, EfficientCSI_1, EfficientCSI_2, CRNet}. As the first CSI feedback network proposed in \cite{csinet}, CsiNet introduced a convolutional neural network (CNN)-based auto-encoder framework to encode high-dimensional CSI into a lower-dimensional latent space, achieving overwhelming superiority over conventional methods. Building on this success, various extensions have been proposed to address time-varying channels \cite{CsiNet_ext1}, reconfigurable intelligent surfaces (RIS) \cite{starris}, and scenario-adaptive architectures \cite{autocsinet}. In addition to scenario extensions, CsiNet-LSTM \cite{CsiNet_ext1}, CsiNet+ \cite{CsiNet+}, and Attention-CSI \cite{Attention-CSI} incorporated advanced modules—recurrent units, refined convolutional kernels, and attention mechanisms—to further enhance performance. With the advent of Transformer architectures \cite{transformer}, recent works \cite{TransNet1, TransNet2} also leveraged their ability to model long-range dependencies in CSI data. In parallel with these architectural improvements, UpAid-FBnet \cite{FBnet} and DualNet \cite{DualNet} exploited uplink instantaneous CSI to assist CSI reconstruction at the BS by leveraging partial reciprocity in FDD systems. Building upon this, CovNet \cite{CovNet} explicitly incorporated uplink covariance to further improve CSI feedback and reconstruction performance, especially under high compression ratios.
	
	Beyond conventional DL-based solutions, large AI models (LAMs) have attracted increasing interest in the wireless community \cite{gpt}. By integrating massive pretrained knowledge with advanced reasoning, contextual understanding, and superior generalization, LAMs offer a new paradigm for intelligent wireless system design. For example, multi-agent systems have been proposed to optimize 6G networks using natural language instructions, combined with data retrieval, planning, and evaluation \cite{multiagent}. Similarly, \cite{LAMSC} presented a semantic communication framework for image transmission, incorporating segmentation-based knowledge construction, attention integration, and adaptive compression. In the context of CSI feedback, large language models (LLMs) have also demonstrated promising performance gains \cite{LLM4CF}. However, the real-time deployment of LAMs remains constrained by their high computational and memory demands, as well as inference latency. This highlights the need to reconcile the expressiveness of LAMs and the stringent deployment constraints of practical wireless systems.
	
	While prior DL-based approaches have achieved remarkable success in improving CSI feedback and reconstruction accuracy, their practical deployment remains hindered by the substantial inference latency and computational overhead associated with real-time processing. To overcome this challenge, recent work has explored DL-aided codebook optimization \cite{AI4C2F}, shifting the learning burden offline and eliminating the need for real-time inference. However, such methods often employ lightweight models with limited representational capacity, which constrains their ability to model complex CSI structures and generalize across diverse environments. Therefore, it is crucial to revisit the codebook-based CSI feedback—not merely as a legacy solution, but as a standard-compatible and inference-free framework that can be significantly enhanced by integrating more expressive learning models. The above overview raises a critical question: \textit{How can LAMs be integrated into CSI feedback to achieve outstanding trade-off between advanced learning capabilities and practical deployment constraints?} 
	
	To answer this question, we propose two novel frameworks leveraging LAM, specifically a large vision model (LVM) pre-trained on large-scale image datasets, to enhance CSI feedback performance via offline codebook optimization. By leveraging LAMs to optimize codebooks offline while avoiding online inference overhead, the proposed frameworks embody the principle of \textit{``Extract the Best, Discard the Rest"}. In summary, our major contributions are as follows.

	\begin{itemize}
		\item \textbf{Site-Specific LAM-enhanced CSI Feedback Framework:} We propose the site-specific LAM-enhanced CSI feedback (SSLCF) framework to provide customized CSI codebooks tailored to specific BS deployments. SSLCF fine-tunes a pretrained LAM using site-specific CSI data to generate an enhanced codebook that reflects local propagation characteristics. It supports both single-side (SS) deployments, where the UE retains the conventional RVQ codebook, and dual-side (DS) deployments, where the enhanced codebook is synchronized to both ends for coordinated feedback.
		\item \textbf{Multi-Scenario LAM-enhanced CSI Feedback Framework:} To improve generalization across diverse deployment environments, we further propose the multi-scenario LAM-enhanced CSI feedback (MSLCF) framework. MSLCF is exclusively designed for dual-side deployments, where both the BS and UE locally store a shared set of enhanced codebooks, each fine-tuned for a specific environment type. Once deployed, the BS identifies its environment type and transmits the corresponding codebook index to the UE, enabling lightweight codebook selection without full-codebook transmission or retraining.
		\item \textbf{Network and Training Design for LAM-Based Codebook Optimization:} To support
		the proposed CSI feedback frameworks, we develop a codebook optimization network named LVM4CF, which adapts a LLaMA-based vision backbone for refining codewords from the conventional RVQ codebook. Building on this network, we propose two offline training strategies tailored to different deployment requirements. For single-side deployments, we train the model to enhance the codebook only at the BS while maintaining compatibility with the UE-side RVQ codebook. For dual-side deployments, we incorporate a codebook update mechanism that iteratively improves the shared codebook used by both the BS and the UE, enabling end-to-end coordination.
	\end{itemize}
	
		This paper is structured as follows: Section~\ref{Section 2} introduces the system model and conventional CSI feedback frameworks. Section~\ref{Section 3} proposes the offline LAM enhanced CSI feedback framework, including the SSLCF and MSLCF schemes. Section~\ref{Section 4} describes the LVM4CF architecture and training algorithm. Simulation results are provided in Section~\ref{Section 5}, and conclusions of this paper are drawn in Section~\ref{Section 6}.
	
\textbf{Notation:} $x, \mathbf{x}, \mathbf{X}$ represent scalar, column vector, and matrix, respectively. $(\cdot)^T, \ (\cdot)^H, \ (\cdot)^{-1}$ respectively denote the transpose, transpose-conjugate, and inverse operations. $\|\cdot\|_2$ denotes the $\ell_2$-norm. The operator $\mathbb{E}\{\cdot\}$ denotes the expectation. $\mathrm{Re}\{\cdot\}$ and $\mathrm{Im}\{\cdot\}$ denote the real and imaginary parts of a scalar, vector, or matrix.
${\bf x}_{[i:j]}$ denote the $i$-th to $j$-th elements of ${\bf x}$, respectively. The expression $\mathcal{CN}(\mu,\sigma^2)$ denotes a circularly symmetric Gaussian distribution with mean $\mu$ and variance $\sigma^2$. $\mathbb{C}^{M \times N}$ and $\mathbb{R}^{M \times N}$ represent the set of $M \times N$ dimension real- and
complex-valued matrixes. $k \in \mathcal{K}$ means the element $k$ belongs to the set $\mathcal{K}$.
	
\section{System Model and CSI Feedback}\label{Section 2}
	
\subsection{System Model}\label{system model}
	Consider an FDD massive MIMO system where a BS with $N_{\rm T}$ antennas transmits signals to $K$ UEs, each equipped with a single omnidirectional antenna. The BS employs a dual-polarized uniform planar array (UPA) with $N_{\rm H}$ horizontal and $N_{\rm V}$ vertical elements, resulting in $N_{\rm T} = N_{\rm H} \times N_{\rm V}$ and $N_{\rm c} = 2N_{\rm T}$ total antenna elements. The channel to the $k$-th UE is denoted as ${\bf h}_k\in\mathbb{C}^{N_{\rm c}\times 1}$. With precoding technologies, the received signal of the $k$-th UE can be expressed as
	\begin{equation}
		y_k = {\bf h}_k^H {\bf v}_k s_k + \sum_{j \ne k} {\bf h}_k^H {\bf v}_j s_j + n_k,
	\end{equation}
	where $y_k$ is the received signal for the $k$-th UE, $s_k$ is the transmitted symbol for the $k$-th UE, and $n_k$ denotes the additive noise following $\mathcal{CN}(0,\sigma^2)$ at the $k$-th UE. The term ${\bf v}_k s_k$ represents the desired signal component, while $\sum_{j \ne k} {\bf h}_k^H {\bf v}_j s_j$ corresponds to the interference from other UEs. ${\bf V} = [{\bf v}_1, {\bf v}_2, \ldots, {\bf v}_k] \in \mathbb{C}^{N_{\rm c} \times K}$ is the linear precoding matrix designed based on the estimated CSI ${\hat {\bf H}}$, where ${\hat {\bf H}} = [{\hat {\bf h}}_1, {\hat {\bf h}}_2, \cdots, {\hat {\bf h}}_K]^T \in \mathbb{C}^{K \times N_{\rm c}}$. In particular, the precoding matrix ${\bf V}$ can be obtained by
	\begin{align}
		{\bf V} = {\rm Prec}({{\hat {\bf H}}}, \sigma^2),
	\end{align}
	where ${\rm Prec}(\cdot)$ represents the precoding scheme, e.g., zero forcing (ZF) \cite{ZF}, minimum mean square error (MMSE) \cite{MMSE}, and weighted MMSE (WMMSE) \cite{WMMSE}. Taking the widely used ZF precoding as an example, the precoding matrix is given by
	\begin{equation}
		\mathbf{V}_{\mathrm{ZF}} = c \hat {\bf H}^H \left(\hat {\bf H} \hat {\bf H}^H\right)^{-1},
	\end{equation}
	where $c$ denotes the power normalization factor defined as
    \begin{equation}
        c = \sqrt{\frac{P}{\|\hat {\bf H}^H(\hat {\bf H} \hat {\bf H}^H)^{-1}\|_F^2}},
    \end{equation}
    and $P$ denotes the total transmit power available at BS. The sum rate is given by \cite{TENN}
	\begin{equation}
		R = \sum_{k=1}^{K}\log_2\left(1 + \frac{|\mathbf{h}_k^H \mathbf{v}_k|^2}{\sum_{i \neq k} |\mathbf{h}_k^H \mathbf{v}_i|^2 + \sigma^2}\right).
	\end{equation}

\subsection{Codebook-Based CSI Feedback}

In FDD systems, the downlink CSI is estimated at the UE side and fed back to the BS, where it is subsequently adopted to enhance downlink transmission, such as precoder design. Most existing DL-based CSI feedback schemes focus on end-to-end CSI compression and reconstruction frameworks from the BS to the UE \cite{DL4CF_overview}. However, such approaches have several challenges, including increased computational burden at both BS and UE and high inference latency. In contrast, we consider the codebook-based CSI feedback \cite{CB4CF_overview, AI4C2F}. Specifically, the UE selects the codeword from a predefined codebook ${\mathcal C}_{\rm u}$ that best matches the channel (assuming perfect CSI estimation at the UE side), and obtains its corresponding codeword index
\begin{align}
	i_k \!=\! \arg\max_{1\leq i\leq 2^B} \left|\mathbf{h}_k^H\,{\bf c}_{i,\mathrm{u}}\right|,\ {\mathcal C}_{\rm u} = \{{\bf c}_{1,\mathrm{u}}, {\bf c}_{2,\mathrm{u}}, \ldots, {\bf c}_{2^B,\mathrm{u}}\},
\end{align}
where ${\bf c}_i\in\mathbb{C}^{N_{\rm c}\times 1}$ is the unit-norm codeword, and $B$ is the number of feedback bits. Subsequently, the UE selects the index $i_k$ and feeds it back to the BS. The BS then selects the corresponding codeword from the predefined codebook as the estimated CSI, i.e., 
\begin{align}
	{\hat {\bf h}}_k = {\bf c}_{i_k,{\rm b}},\ {\mathcal C}_{{\rm b}} = \{{\bf c}_{1,{\rm b}}, {\bf c}_{2,{\rm b}}, \ldots, {\bf c}_{2^B,{\rm b}}\}.
\end{align}
Note that in the conventional schemes, the same codebook is typically deployed at both the BS and UE sides, i.e., ${\mathcal C}_{\rm u} = {\mathcal C}_{\rm b} = {\mathcal C}$, which is designed without considering environment-aware characteristics.

    \begin{figure}[!t]
			\centering
			\includegraphics[width = \linewidth]{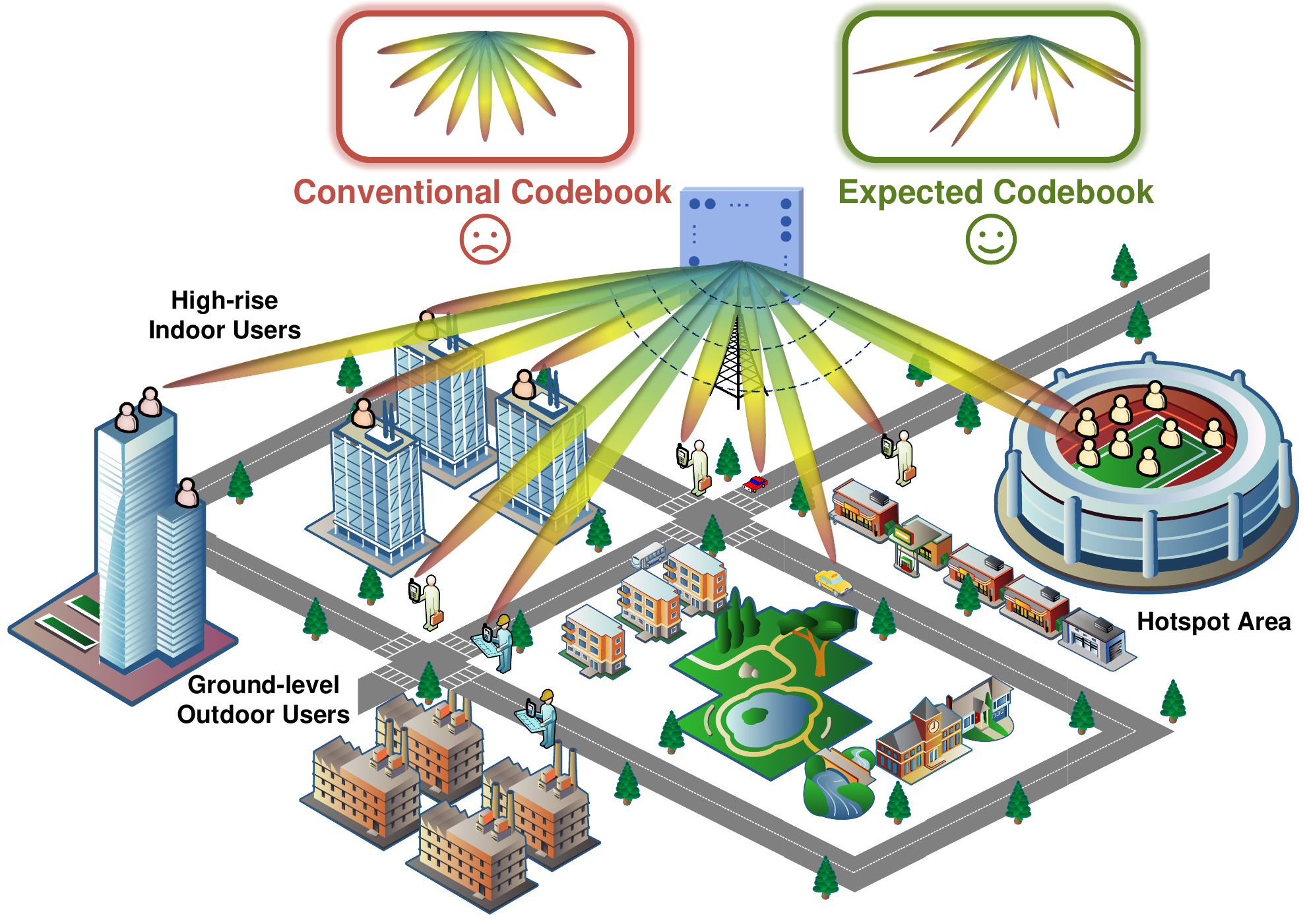}
			\caption{Example of deployment scenario with conventional and environment-aware codebook.}
			\label{fig:sys_model}
	\end{figure}	
	
	\begin{figure*}[!t]
		\centering
		\subfigure[Site-specific LAM-enhanced CSI feedback framework.]{
			\begin{minipage}[t]{0.48\linewidth}
				\centering
				\includegraphics[width=0.85\textwidth]{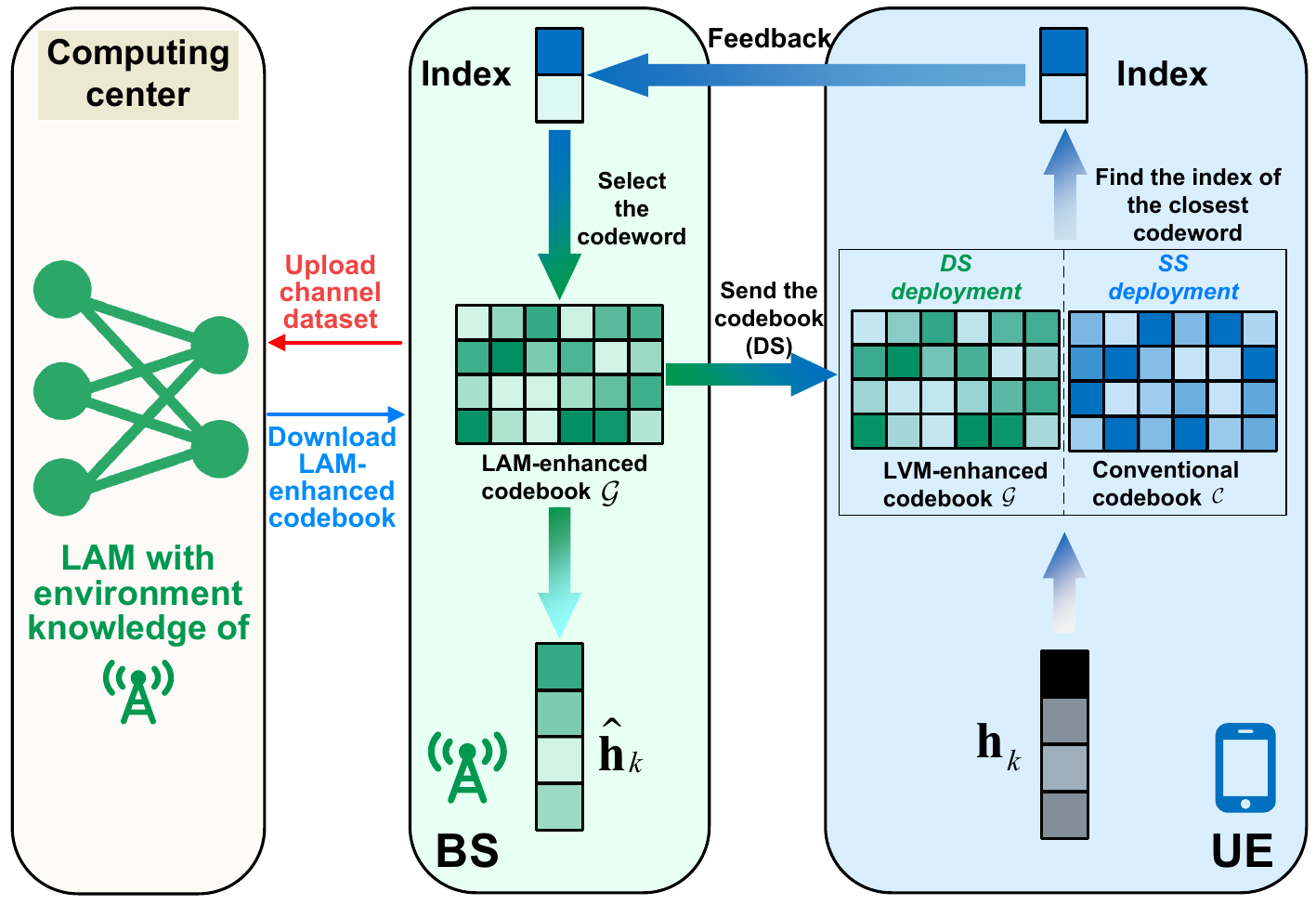}
			\end{minipage}%
		}%
		\subfigure[Multi-scenario LAM-enhanced CSI feedback framework.]{
			\begin{minipage}[t]{0.48\linewidth}
				\centering
				\includegraphics[width=1.0\textwidth]{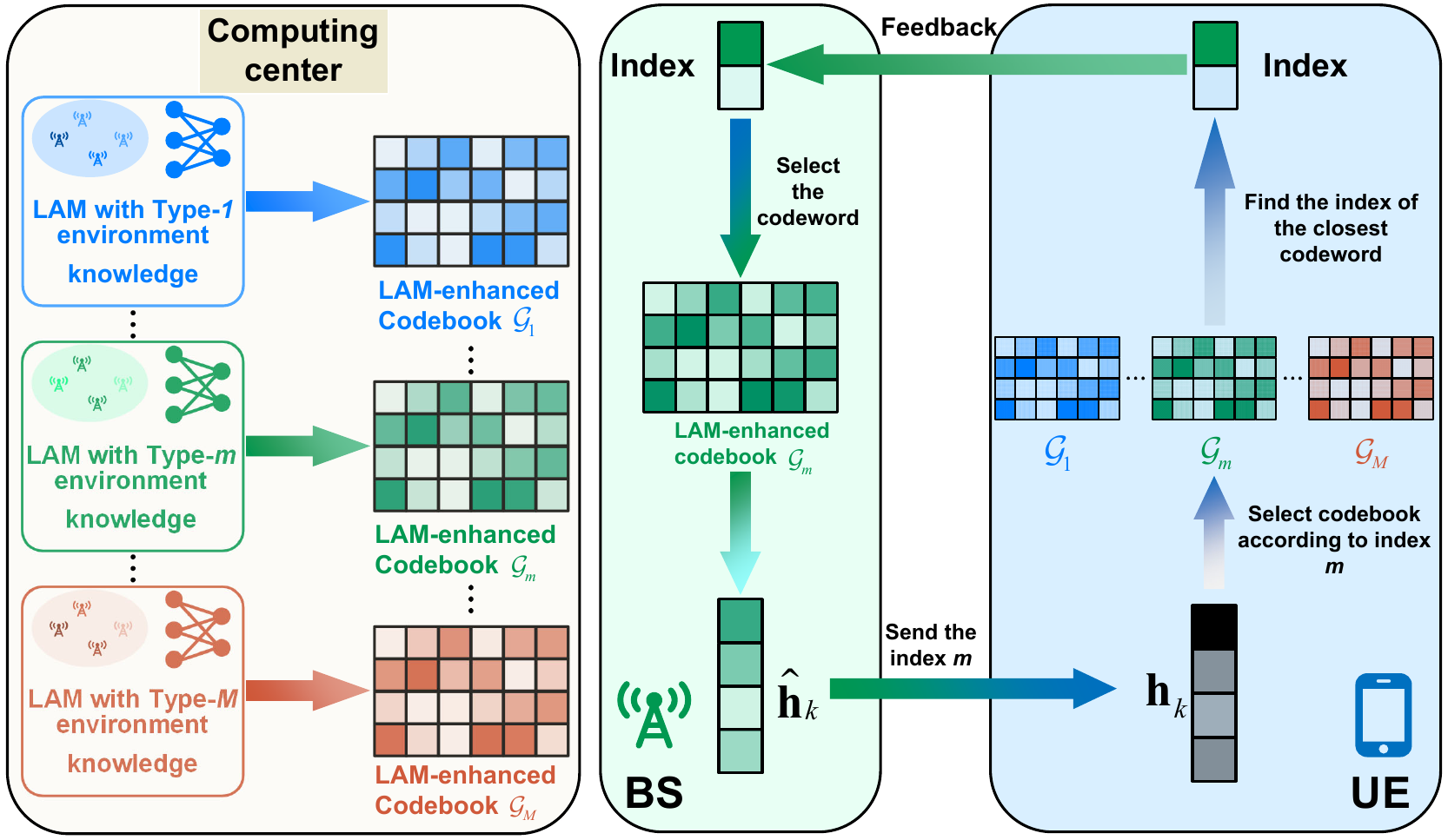}
			\end{minipage}%
		}%
		\centering
		\caption{The two proposed codebook-based CSI feedback frameworks.}
		\label{fig:Framework}
	\end{figure*}
	
	\subsection{Problem Formulation}\label{problem formulation}
	
Although codebook-based schemes offer advantages such as simple protocol design, low implementation complexity, and ease of standardization \cite{3gpp_ts_38_214}, they inherently suffer from quantization errors, particularly in massive MIMO scenarios with high-dimensional CSI \cite{hou2024beam}. Since conventional codebooks are typically constructed via isotropic beam domain sampling without considering site-specific characteristics \cite{sscodebook}, they fail to capture the non-uniform channel distribution in practical deployments, leading to significant quantization errors under limited feedback overhead. 

To address this, environment-aware codebooks are required to align with local channel characteristics. This naturally leads to two deployment options: using the same enhanced codebook at both the BS and UE—referred to as DS deployment, or applying the enhanced codebook solely at the BS while maintaining the conventional one at the UE—referred to as SS deployment.
To formalize the design of environment-aware codebooks, we formulate the following optimization problem based on the underlying channel distribution. Specifically, assuming that the downlink channel follows $\mathbf{h}\sim p(\mathbf{h})$, we aim to optimize the reconstruction codebook $\mathcal{G}=\{\mathbf{g}_{1}, \ldots, \mathbf{g}_{2^B}\}$ as follows
\begin{subequations}\label{eq:problem_SS}
	\begin{align}
		({\rm P1})\quad{\mathcal G}^{\star}=\arg\max_{{\mathcal G}}\;&
		\mathbb{E}_{\mathbf{h}\sim p(\mathbf{h})}
		\left\{
		\frac{\bigl|f^{H}(\mathbf{h}, {\mathcal C}, {\mathcal G})\,\mathbf{h}\bigr|}
		{\|f(\mathbf{h}, {\mathcal C}, {\mathcal G})\|_{2} \|\mathbf{h}\|_{2}}
		\right\}
		\label{eq:prob_SS_obj}\\
		\mathrm{s.t.}\;
		\|\mathbf{g}_i\|_2 & = 1,\quad i=1,\ldots,2^B,
		\label{eq:prob_SS_cons}
	\end{align}
\end{subequations}
where $f(\mathbf{h}, {\mathcal C}, {\mathcal G}) = {\bf g}_i$ denotes the reconstructed channel at the BS using the enhanced codebook $\mathcal{G}$, with the index $i$ determined at the UE side using the conventional codebook $\mathcal{C}$ for SS deployment or enhanced codebook $\mathcal{G}$ for DS deployment, and $i$ is determined by
\begin{align}\label{eq:index_selection}
i=
\begin{cases}
 \arg\max_{1\le j\le2^B}
\frac{\bigl|\mathbf{c}_j^H\,\mathbf{h}\bigr|}
{\|\mathbf{c}_j\|_2\,\|\mathbf{h}\|_2}, \quad {\rm SS}\\
 \arg\max_{1\le j\le2^B}
\frac{\bigl|\mathbf{g}_j^H\,\mathbf{h}\bigr|}
{\|\mathbf{g}_j\|_2\,\|\mathbf{h}\|_2}, \quad {\rm DS}
\end{cases}.
\end{align}
The optimization problem $({\rm P1})$ provides a unified formulation for both the SS and DS deployment modes. Specifically, for \textit{SS deployment}, the UE employs the conventional codebook $\mathcal{C}=\{\mathbf{c}_1,\dots,\mathbf{c}_{2^B}\}$ for quantization, while the BS reconstructs the channel using the enhanced codebook $\mathcal{G} = \{\mathbf{g}_{1}, \ldots, \mathbf{g}_{2^B}\}$. In contrast, \textit{DS deployment} assumes that the enhanced codebook $\mathcal{G}$ is shared by both the UE and BS, i.e., $\mathcal{C} = \mathcal{G}$. 
As shown in \eqref{eq:index_selection}, these two deployment modes lead to different formulations of the function $f$, depending on which codebook is employed for quantization at the UE. This distinction introduces flexibility in deployment while maintaining backward compatibility with conventional schemes, which will be discussed in subsequent sections. In the problem $({\rm P1})$, the optimal design of ${\mathcal G}$ can be interpreted as the mapping $G$: $ {\mathcal G}^{\star} = G\left[\mathcal{C}, p(\mathbf{h})\right]$. 

To intuitively illustrate the impact of codebook improvement in problem $({\rm P1})$, Fig.~\ref{fig:sys_model} depicts a representative deployment scenario where UEs are clustered in ground-level outdoor areas, high-rise indoor regions, and specific hotspots. In such cases, the conventional codebook allocates codewords uniformly across the entire beam domain without accounting for the UE distribution, leading to resource waste in sparse regions and insufficient resolution in dense regions, thereby hindering the accurate capture of dominant channel directions. To address this mismatch, the desirable environment-aware codebook would be site-specific and allocate codewords more adaptively, with denser sampling in high-density regions and vice versa in low-density regions, thereby better aligning with the spatial characteristics of given deployments and improving CSI reconstruction accuracy.  

Although codebook enhancement is expected to improve the performance of CSI feedback, it remains highly challenging to solve problem $({\rm P1})$. On the one hand, the underlying channel distribution $p(\mathbf{h})$ is typically unknown and lacks an explicit analytical form, making the expectation in the objective function intractable. On the other hand, the reconstruction function \(f\) involves discrete index selection, which is inherently non-differentiable and thus incompatible with most gradient-based optimization techniques.
	
\section{Offline Large AI Model-Enhanced \\CSI Feedback Framework}\label{Section 3}

Given the challenges of problem $({\rm P1})$, we depart from traditional optimization techniques and instead directly learn the mapping from the channel distribution and conventional codebook to the enhanced codebook, i.e., ${\mathcal G}^{\star} = G\left[\mathcal{C}, p(\mathbf{h})\right]$.
To effectively capture site-specific characteristics inherently present in the channels, we propose two LAM-assisted CSI feedback frameworks, namely SSLCF and MSLCF, which leverage the strong representational power of LAM to learn this mapping.
Specifically, the SSLCF framework proposed in \secref{Section3A} generates site-specific enhanced codebooks for CSI feedback, closely matching the channel characteristics at each BS deployment. In contrast, the MSLCF framework proposed in \secref{Section3B} produces multiple scenario-specific enhanced codebooks by fine-tuning the pre-trained LAM using diverse CSI datasets collected from multiple BS environments. 
    
    Both frameworks consist of two stages: an \emph{offline codebook generation stage}, where the LAM is adapted to the target BS environment or multiple BS environments, then used to generate enhanced codebooks; and an \emph{online CSI feedback stage}, where the enhanced codebooks are used at the BS for high-quality CSI reconstruction in both SS and DS deployments, and at the UE for accurate CSI quantization in DS deployments. The proposed frameworks are illustrated in Fig.~\ref{fig:Framework}. We refer to this design principle as ``\textbf{Extract the Best, Discard the Rest}''. Specifically, The strong expressive power of the LAM is exploited to enhance the performance of CSI feedback, corresponding to \textbf{Extract the Best}. Meanwhile, offline training and inference avoid the long latency and deployment challenges of large models at the BS or UE side, corresponding to \textbf{Discard the Rest}.
	
	\subsection{Site-Specific LAM-enhanced CSI Feedback Framework}\label{Section3A}

    As previous analysis suggests, conventional codebooks for CSI feedback that are isotropic in the beam domain limit their the ability to adapt to the specific propagation characteristics of each deployment site, which in practice often leads to suboptimal performance. However, once the BS is deployed, the wireless propagation environment within its coverage area tends to remain statistically stable over relatively long timescales. This observation motivates the development of site-specific enhanced codebooks that more accurately capture the local channel characteristics. In the proposed SSLCF framework, each BS fine-tunes a pretrained LAM on locally collected CSI data, resulting in a customized codebook optimized for the statistical properties of its deployment environment.


	\subsubsection{Offline Stage}
	In the offline stage, the BS first collects a local channel dataset, denoted as $\mathcal{H}_{\mathrm{train}}$. This dataset is then uploaded to a computing center (e.g., via fiber-optic links). At the computing center, a LAM that has been pre-trained on large-scale datasets (e.g., text, image, or wireless channel datasets \cite{LWM}) is fine-tuned using $\mathcal{H}_{\mathrm{train}}$, allowing the model to adapt to the site-specific propagation characteristics surrounding the BS.
	
	Once fine-tuned, the LAM is used to enhance the conventional codebook ${\mathcal C} = \{{\bf c}_1, {\bf c}_2, ..., {\bf c}_{2^B}\}$. Specifically, for each codeword ${\bf c}_i$ in ${\mathcal C}$, the LAM outputs a refined version, which is given by
	\begin{align}
		{\bf g}_i = f_{\text{LAM}}({\bf c}_i; \Theta),\quad \forall i \in \{1, ..., 2^B\},
	\end{align}
	where $\Theta$ denotes the LAM parameters learned during fine-tuning. The resulting enhanced codebook $\mathcal{G} = \{{\bf g}_1, {\bf g}_2, \ldots, {\bf g}_{2^B}\}$ is then downloaded by the BS and stored locally for use in CSI reconstruction.

	\subsubsection{Online Stage}
	In the online stage, codebook-based CSI feedback is executed. Deploying the enhanced codebook at both the BS and the UE requires the BS to transmit the entire codebook to the UE, incurring substantial overhead. To strike a balance between performance and overhead, SSLCF flexibly supports both SS and DS deployments. In SS deployment, only the BS uses the enhanced codebook $\mathcal{G}$ for CSI reconstruction while the UE continues to use the conventional codebook $\mathcal{C}$, i.e., $\mathcal{C}_{\rm u}=\mathcal{C}$. In DS deployment, the BS transmits $\mathcal{G}$ to the UE via radio resource control (RRC) signaling so that both sides can employ the enhanced codebook for CSI feedback \cite{3gpp_ts_38_331}. At the UE, for each channel vector $\mathbf{h}_k$, the optimal codeword index is computed by maximizing the cosine similarity between $\mathbf{h}_k$ and the candidate codewords from $\mathcal{C}_{\rm u}$. The UE selects the best codeword by
	\begin{align}
    i_{k} = 
    \begin{cases}
     \arg\max_{1\le j\le2^B}
    \frac{\bigl|\mathbf{c}_j^H\,\mathbf{h}\bigr|}
    {\|\mathbf{c}_j\|_2\,\|\mathbf{h}\|_2}, & \text{SS deployment}\\
     \arg\max_{1\le j\le2^B}
    \frac{\bigl|\mathbf{g}_j^H\,\mathbf{h}\bigr|}
    {\|\mathbf{g}_j\|_2\,\|\mathbf{h}\|_2}, & \text{DS deployment}
    \end{cases}.
    \end{align}
    and the index $i_k$ is fed back to the BS and subsequently adopted to retrieve ${\bf g}_{i_k}$ as an estimate of $\mathbf{h}_k = {\bf g}_{i_k}$ based on the enhanced codebook ${\mathcal C}_{{\rm b}} = \mathcal{G} = \{{\bf g}_1, {\bf g}_2, \ldots, {\bf g}_{2^B}\}$.

	\subsection{Multi-Scenario LAM-enhanced CSI Feedback Framework}\label{Section3B}

    While the SSLCF framework achieves excellent CSI feedback performance by generating highly customized codebooks fine-tuned with site-specific CSI data, scaling it to large-scale or dynamic deployments poses additional challenges, particularly in terms of signaling overhead during dual-side deployments. To address these evolving needs, the MSLCF framework generates a set of environment-aware codebooks by fine-tuning offline with diverse CSI datasets from multiple representative environments, enabling the system to reduce signaling overhead by transmitting only a compact environment index along with the codeword index. By doing so, the MSLCF framework can achieve efficient and scalable CSI feedback across heterogeneous scenarios.
    	
	\subsubsection{Offline Stage}
	For each environment type $m \in \mathcal{M}$, the LAM is fine-tuned on the corresponding CSI data, i.e., $\mathcal{H}_{\mathrm{train},m}$, which leads the model adapting to the specific type of BS environment, e.g., urban macro (UMa) and rural macro (RMa) environments. The fine-tuned LAM then refines the codeword in $\mathcal{C}$. Specifically, for every codeword $\mathbf{c}_i \in \mathcal{C}$, the LAM tuned for environment $m$ generates an enhanced codeword, given by
	\begin{equation}
		\mathbf{g}_{m,i} = f_{\text{LAM}, m}(\mathbf{c}_i; \Theta_m), \ \forall m \in \mathcal{M},\ \forall i \in \{1,\dots,2^B\}
	\end{equation}
	where $\Theta_m$ denotes the parameters optimized for environment type $m$. The enhanced codebook for environment $m$ is then given by
	\begin{equation}
		\mathcal{G}_m = \{\mathbf{g}_{1,m}, \mathbf{g}_{2,m}, \ldots, \mathbf{g}_{2^B,m}\}.
	\end{equation}
	After the offline generation, the set of enhanced codebooks $\{\mathcal{G}_m\}_{m \in \mathcal{M}}$ is deployed at the UE, while the BS retains the enhanced codebook corresponding to its own environment.
	
	\subsubsection{Online Stage}
	Prior to online feedback, the BS determines its environment type $m \in \mathcal{M}$ based on its hardware configuration and deployment geography. Then, the BS transmits an environment index to the UE, instructing it to select the appropriate codebook. Because only an environment index—rather than a full codebook—must be delivered, MSLCF inherently benefits from this low overhead and therefore adopts DS deployment. Upon receiving this indicator, the UE selects the corresponding enhanced codebook $\mathcal{G}_m$. Therefore, both BS and UE deploy the same enhanced codebook, given by
	\begin{equation}
	\mathcal{C}_{\rm u}=\mathcal{C}_{\rm b}=\mathcal{G}_m.
	\end{equation}
	Then, for each CSI vector $\mathbf{h}_k$, the UE computes the optimal codeword index by maximizing the cosine similarity, which can be expressed as
	\begin{equation}
		i_k = \arg\max_{1 \le i \le 2^B} \frac{\bigl|\mathbf{g}_{i,m}^H\,\mathbf{h}_k\bigr|}{\|\mathbf{g}_{i,m}\|_2\,\|\mathbf{h}_k\|_2}.
	\end{equation}
	The UE feeds back the index $i_k$ to the BS, which then performs a lookup in the enhanced codebook $\mathcal{G}_m$ to retrieve the corresponding codeword for CSI reconstruction.

	\subsection{Comparison Between SSLCF and MSLCF Frameworks}
	The differences between the two frameworks can be summarized as follows:
	
	\begin{itemize}
		\item \textbf{Deployment Mode:} In SSLCF, the BS uploads its channel dataset to the computing center, which returns a site-specific enhanced codebook, requiring periodic data exchange for updates. In contrast, MSLCF pre-generates multiple enhanced codebooks tailored to different environmental conditions, eliminating the need for real-time interaction between the BS and the computing center.
		
		\item \textbf{Signaling Overhead:} In SSLCF, if choosing the DS deployment, the entire enhanced codebook must be delivered to the UE, incurring higher overhead, while MSLCF only requires the BS to signal a compact environment indicator to select the appropriate pre-deployed codebook.
		
		\item \textbf{Adaptation Granularity:} SSLCF produces a highly adapted codebook specific to a single BS environment, but MSLCF generates a generalized codebook for each environment type, trading some precision for broader applicability.
		
	\end{itemize}

\begin{figure*}[t]
\centering
\includegraphics[width=6.5in]{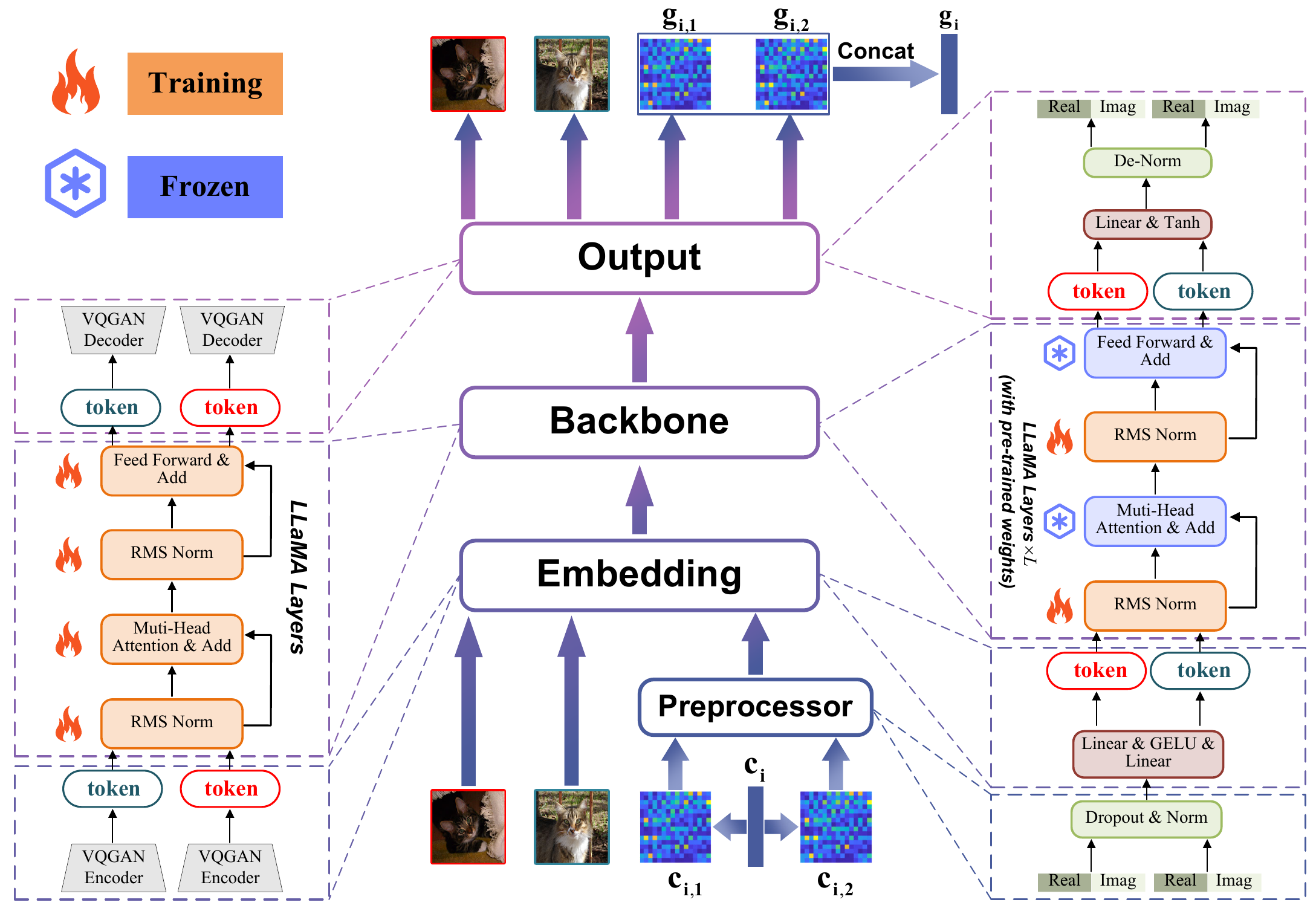}
\caption{Network architectures of LVM4CF: Vision task vs. CSI codebook enhancement.}
\label{LVM4CF}	
\end{figure*}

	\section{LVM4CF: Customized LVM for CSI Feedback}\label{Section 4}
	
	In this section, we propose network ``LVM4CF'' to serve as the function $f_{\rm LAM}$ in the two proposed frameworks. LVM4CF leverages a LLaMA-based vision model to refine CSI codewords from the conventional codebook, thereby converting quantized CSI codewords into enhanced representations that more accurately reflect the actual CSI. This design, which combines large-scale visual pre-training with subsequent fine-tuning on CSI data to capture the intricate characteristics of the channel, significantly improves the performance of codebook-based CSI feedback.

	\subsection{Network Architecture of LVM4CF}\label{network} 
	As shown in \figref{LVM4CF}, LVM4CF integrates a preprocessing module that formats CSI data, an embedding module that projects the data into a high-dimensional latent space, a backbone network (adapted from LLaMA) that extracts features, and an output layer that reconstructs the refined CSI codewords. 
	
	\subsubsection{Preprocessing} Since the BS employs dual‐polarized antennas, each CSI codeword from the conventional codebook has a dimension $N_c = 2N_{\rm T}$, which can be partitioned into two equal-length subvectors corresponding to the two polarizations $p\in\{1, 2\}$. Each codeword $ {\bf c}_i \in \mathbb{C}^{N_c \times 1}$ is divided into two subvectors as
	\begin{align}
		{\bf c}_{i,1} = {\bf c}_{i[1:N_{\rm T}]},\ {\bf c}_{i,2} = {\bf c}_{i[(N_{{\rm T}}+1):2N_{\rm T}]}.
	\end{align}
	The real and imaginary parts of the above vector are arranged into the input matrix, expressed by
	\begin{align}
		{\bf X} = [{\bf x}_{\mathrm{in},1},{\bf x}_{\mathrm{in},2}]^T,\ 	{\bf x}_{\mathrm{in},p} = \begin{bmatrix} \Re\{{\bf c}_{i,p}\} \\ \Im\{{\bf c}_{i,p}\}
		\end{bmatrix}, p\in\{1, 2\}.
	\end{align}
	\subsubsection{Embedding} In this stage, each token is normalized and projected into a high-dimensional latent space via a token-wise feedforward network, given by
	\begin{align}
		{\bf x}_{{\rm e},p} \!=\! \mathbf{W}_2\Bigl(\mathrm{GELU}\bigl(\mathrm{LN}(\mathbf{W}_1\,{\bf x}_{\mathrm{in},p} \!+\! \mathbf{b}_1)\bigr)\Bigr) \!+\! \mathbf{b}_2,
	\end{align}
	where $\mathbf{W}_1\in\mathbb{R}^{(D/2) \times N_c}$, $\mathbf{W}_2\in\mathbb{R}^{D\times (D/2)}$, and $D$ is the hidden layer dimension of the LLaMA model. $\mathrm{GELU}$ denotes the Gaussian error linear unit activation function \cite{GELU} and $\mathrm{LN}$ denotes layer normalization \cite{LN}. The weight matrices $\mathbf{W}_1$, $\mathbf{W}_2$ and bias vectors $\mathbf{b}_1$, $\mathbf{b}_2$ are learnable parameters.
	The embedded tokens form the sequence as
	\begin{align}
		{\bf X}_{\rm e} = [{\bf x}_{\mathrm{e},1},{\bf x}_{\mathrm{e},2}]^T.
	\end{align}
	Then, rotary position embedding \cite{RoPE} is applied to ${\bf X}_{\rm e}$ to incorporate positional information.
	
	\subsubsection{Backbone} The backbone consists of $L$ transformer layers. Let ${\bf X}^{(0)} = {\bf X}_e$ denote the input to the first layer. The $l$-th ($1\le l\le L$) layer first computes an attention-enhanced representation and then refines it via a feedforward network, which can be expressed as
	\begin{align}
		\begin{aligned}
			{\bf X}^{(l)}_{\mathrm{att}} &= {\bf X}^{(l-1)} + \mathrm{MA}\bigl(\mathrm{RMSNorm}({\bf X}^{(l-1)})\bigr), \\
			{\bf X}^{(l)} &= {\bf X}^{(l)}_{\mathrm{att}} + \mathrm{FFN}\bigl(\mathrm{RMSNorm}({\bf X}^{(l)}_{\mathrm{att}})\bigr).
		\end{aligned}
	\end{align}
	$\mathrm{MA}$, $\mathrm{RMSNorm}$, and $\mathrm{FFN}$ denotes multi-head attention block \cite{transformer}, root mean square normalization \cite{RMSNorm}, and feed-forward network, respectively.
	The output of the final layer is ${\bf X}_{\rm LVM} = {\bf X}^{(L)} = [{\bf x}_{{\rm LVM},1}, {\bf x}_{{\rm LVM},2}]$,
	which contains two tokens, each being a $d_{\rm LVM}$-dimensional vector.
	
	\subsubsection{Output} In this stage, each token is mapped back to a real-valued vector of dimension $N_c$ using a fully connected (FC) layer with activation function, given by
	\begin{align}
		{\bf x}_{\mathrm{out},p} = {\mathrm{Tanh}}\bigl(\mathbf{W}_3\,{\bf x}_{{\rm LVM},p} + \mathbf{b}_3\bigr),\quad p\in\{1, 2\},
	\end{align}
	where $\mathbf{W}_3\in\mathbb{R}^{N_c\times D}$. ${\mathrm{Tanh}}$ denotes the hyperbolic tangent activation function \cite{Tanh}, and the weight matrix $\mathbf{W}_3$ along with the bias vectors $\mathbf{b}_3$ are learnable parameters used to map the latent representations back to the output space.
	Finally, the new codeword ${\bf g}_{i}$ is reconstructed as ${\bf g}_{i}= [ \mathbf{g}_{i,1}^{T},  \mathbf{g}_{i,2}^{T}]^{T}$, where we have  
	\begin{align}
		{\bf g}_{i,p} \!=\! {\bf x}_{\mathrm{out},p[1:N_{\rm T}]} + j{\bf x}_{\mathrm{out},p[(N_{{\rm T}}+1):2N_{\rm T}]}, \ \ p\in\{1, 2\},
	\end{align}
	By partitioning the CSI vector into two subvectors according
	to its inherent dual-polarized structure, the network can better
	exploit the complementary features of each polarization, thus
	enhancing the quality of CSI reconstruction.

	\subsection{Backbone Pre-Training and Adaptation for CSI Feedback}
	
	The LVM4CF is pre-trained with a two-stage approach that transforms raw images into a sequence of discrete tokens and then learns the sequential dependencies among these tokens via an autoregressive Transformer, which can be found in \cite{sec3_2}. First, a VQGAN-based \cite{VQGAN} tokenizer is applied to each image, mapping it into a fixed-length sequence of visual tokens. For each specific image, the tokenizer produces a token sequence, expressed by
	\begin{align}
		\mathcal{T} = \{t_1,\, t_2,\, \dots,\, t_n\},
	\end{align}
	where each token \(t_i\) is drawn from a codebook \(\mathcal{K}\) (with \(|\mathcal{K}| = 8192\)) and \(n\) is typically set to 256. The autoregressive Transformer is then trained to predict the next token in the sequence using a standard cross-entropy loss as
	\begin{align}
		\textstyle L= - \sum_{i=1}^{n} \log P\bigl(t_i \mid t_1, t_2, \ldots, t_{i-1}\bigr),
	\end{align}
	This sequential modeling approach enables the model to capture spatial and contextual relationships within images. The VQGAN tokenizer converts high-dimensional visual data into a lower-dimensional discrete representation, and the Transformer learns rich dependencies among these tokens.
	
	The alignment between the vision task and CSI feedback on LVM stems from their shared requirement to model spatial correlations. In LVM's image processing pipeline, the tokenization process decomposes visual data into sequential data, where each token captures local geometric patterns through the VQGAN module. Similarly, the dual-polarized structure of CSI codewords reflects spatial correlation patterns induced by the antenna geometry. By decomposing each codeword into polarization-specific subvectors, we impose a spatial structure on the representation, loosely analogous to image patches. This structural analogy enables the pre-trained LVM to transfer its spatial modeling capability from visual domains to CSI reconstruction tasks.
	
		\begin{algorithm}[t]
\caption{Training and Inference Algorithm for SS Deployment}
	\label{alg:BCT}
	\begin{algorithmic}[1]
		\STATE \textbf{Input:} True channel vectors $\{\mathbf{h}^{(n)}\}_{n=1}^{N}$, baseline codebook ${\mathcal C}$, hyperparameters (batch size, number of epochs $E$, learning rate, decay rate)
		\STATE \textbf{Construct Training Set:} For each ${\bf h}^{(n)}$, select 
		\[ \textstyle
		{\bf c}^{(n)} = \arg\max_{{\bf c} \in {\mathcal C}} \frac{|\mathbf{c}^H\,{\bf h}^{(n)}|}{\|\mathbf{c}\|_2\,\|{\bf h}^{(n)}\|_2}, 
		\]
		and form the training pair $({\bf c}^{(n)}, {\bf h}^{(n)})$. Denote the collection as $\mathcal{H}_{\text{train}} = \{({\bf c}^{(n)}, {\bf h}^{(n)})\}_{n=1}^{N}$.
		\STATE \textbf{Training:} 
		\FOR{each epoch $e = 1$ to $E$}
		\FOR{each batch $\mathcal{H}_{\text{batch}}\subseteq \mathcal{H}_{\text{train}}$}
		\STATE ${\bf g} = {\rm LVM}({\bf c}),\ \forall ({\bf c}, {\bf h})\in\mathcal{H}_{\text{batch}}$.
		\STATE Compute loss: $L_{\rho} 	= -\mathbb{E}_{\mathcal{H}_{\text{batch}}}\{
		\frac{|{\bf g}^H \mathbf{h}|}
		{\|{\bf g}\|_2 \,\|\mathbf{h}\|_2}
		\}.$
		\STATE Backpropagate and update model parameters.
		\ENDFOR
		\STATE Update learning rate using the scheduler.
		\ENDFOR
		\STATE \textbf{Inference:} ${\bf g}_i = {\rm LVM}({\bf c}_i), \forall {\bf c}_i\in{\mathcal C}$. ${\mathcal G}=\{{\bf g}_1,  ..., {\bf g}_{2^B}\}$.
	\end{algorithmic}
\end{algorithm}
	\subsection{Training and Inference Algorithm}\label{Section4B}

\begin{figure}[!t]
        \centering
        \includegraphics[width = 0.96\linewidth]{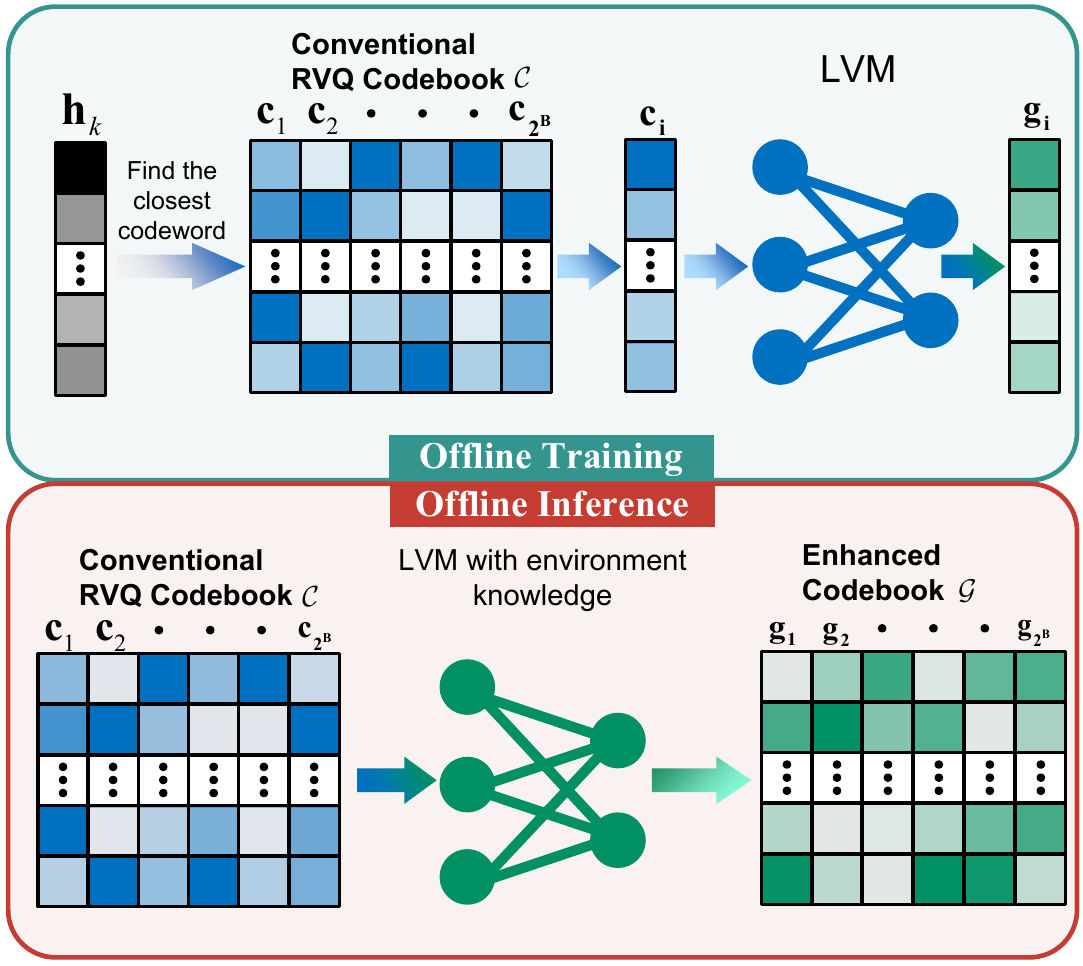}
        \caption{Training strategy for generating enhanced codebook.}
        \label{train_BS}
\end{figure}

In this section, we propose two offline training and inference algorithms that correspond to the two proposed deployment strategies. Both of the two algorithms provide detailed methods for training the network and refine the quantized CSI codewords from the conventional codebook ${\mathcal C}$, denoted by ${\bf c}_i$, to generate an enhanced CSI codeword ${\bf g}_i$.

\subsubsection{Training and inference algorithm for SS deployment} 
		This algorithm generate site-specific enhanced codebooks for SS deployment in SSLCF framework.
		The overall training and inference process is summarized in \algref{alg:BCT}.
		\figref{train_BS} shows a schematic overview of the training strategy. Specifically, in the offline training stage, for each true channel vector, the closest matching codeword from the conventional codebook is selected based on maximum cosine similarity, forming a training pair. The model then refines these selected codewords through supervised learning by minimizing a cosine similarity loss, which optimizes the refined codewords to align more closely with the true channel directions.
		
		The loss function is defined based on the cosine similarity, expressed by
        \begin{align}
        	L_{\rho} = -\mathbb{E}_{\mathcal{H}_{\text{batch}}}\left\{
        	\frac{\left|{\bf g}^H \mathbf{h}\right|}
        	{\left\|{\bf g}\right\|_2 \,\left\|\mathbf{h}\right\|_2}
        	\right\},
        	\label{loss}
        \end{align}
    which aims to maximize the directional alignment between the refined codeword $\mathbf{g}_i$ and the corresponding ground-truth channel vector $\mathbf{h}_{k}$. This training procedure utilizes the channel dataset to fine-tune the LVM4CF and generates a site-specific enhanced codebook. 
    \begin{remark}
        Combining problem $\text{(P1)}$ with the above training and inference processes, the performance gain achieved under the SS deployment can be attributed to the following mechanism: for each codeword in the UE-side conventional codebook, the BS leverages the distribution of the actual channels within the associated channel subspace to refine the codeword, thereby improving system performance.
    \end{remark}

\begin{algorithm}[t]
\caption{Training and Inference Algorithm for DS Deployment}
        \label{alg:DCT}
        \begin{algorithmic}[1]
            \STATE \textbf{Input:} True channel vectors $\{\mathbf{h}^{(n)}\}_{n=1}^{N}$, initial DS codebook ${\mathcal{C}_0=\mathcal C}$, hyperparameters (batch size, number of epochs $E$, learning rate, decay rate)
            \STATE \textbf{Construct Training Set:} For each ${\bf h}^{(n)}$, select 
            \[ \textstyle
            {\bf c}^{(n)}_0 = \arg\max_{{\bf {c}_0} \in {\mathcal {C}_0}} \frac{|\mathbf{c}_0^H\,{\bf h}^{(n)}|}{\|\mathbf{c_0}\|_2\,\|{\bf h}^{(n)}\|_2}, 
            \]
            and form the training pair $({\bf c}_0^{(n)}, {\bf h}^{(n)})$. Denote the collection as $\mathcal{H}_{\text{train}} = \{({\bf c}_0^{(n)}, {\bf h}^{(n)})\}_{n=1}^{N}$.
            \STATE \textbf{Training:}
            \FOR{each epoch \(e = 1\) to \(E\)}
            \IF{(\(e\) mod update\_interval \(= 0\))}
            \STATE \textbf{Inference:} ${\bf c}_{i,{\rm{up}}} = {\rm LVM}({\bf c}_{i,0}), \forall {\bf c}_{i,0}\in{\mathcal {C}_0}$. \\${\mathcal C_{\rm{up}}}=\{{\bf c}_{1,{\rm{up}}},  ..., {\bf c}_{2^B,{\rm{up}}}\}$.
            \STATE Evaluate \(\mathcal{C}_{\text{up}}\) on the validation set; if the validation loss decreases, set \(\mathcal{C}_0 = \mathcal{C}_{\text{up}}\).
            \STATE Reinitialize the trainable parameters in ${\rm LVM}$.
            \STATE Construct training set with the updated $\mathcal{C}_0$.
            \ENDIF
            \FOR{each batch $\mathcal{H}_{\text{batch}}\subseteq \mathcal{H}_{\text{train}}$}
            \STATE ${\bf c}_{\rm{up}} = {\rm LVM}({\bf c}_0),\ \forall ({\bf c}_0, {\bf h})\in\mathcal{H}_{\text{batch}}$.
            \STATE Compute loss: $L_{\rho} 	= -\mathbb{E}_{\mathcal{H}_{\text{batch}}}\{
            \frac{|{\bf c}_{\rm{up}}^H \mathbf{h}|}
            {\|{\bf c}_{\rm{up}}\|_2 \,\|\mathbf{h}\|_2}
            \}.$
            \STATE Backpropagate and update model parameters.
            \ENDFOR
            \STATE Update learning rate using the scheduler.
            \ENDFOR
            \STATE \textbf{Output:} DS enhanced codebook \(\mathcal{G} = \mathcal{C}_0\).
        \end{algorithmic}
\end{algorithm}    
\begin{table}[!t]
	\centering
	\normalsize 
	\caption{Simulation Settings and Training Configurations}
		\begin{tabular}{ll}
			\toprule
			\textbf{Parameter}                    & \textbf{Setting} \\ \midrule
			BS antennas number                    & $N_{\rm T}=64$, $N_{\rm H}=N_{\rm V}=8$ \\ 
			BS antenna model                      & 3GPP-3D \\ 
			BS antenna polarization               & Dual-polarized \\ 
			BS antenna height                     & $25$ m \\ 
			UE antenna				              & Omni \\ 
			UE antenna height				      & $1.5$ m \\ 
			Center frequency (mmWave)             & $28$ GHz \\ 
			Center frequency (sub-6G)             & $3.5$ GHz \\ 
			Scenario (UMa)                        & 3GPP\_38.901\_UMa\_NLOS \\ 
			Scenario (RMa)                        & 3GPP\_38.901\_RMa\_NLOS \\
			Batch size                            & $512$ \\ 
			Optimizer                             & Adam  \\ 
			Epochs (SS)                          & $200$ \\ 
			Epochs (DS)                          & $400$ (Update interval is $40$) \\ 
			Learning rate                        & $0.001$ \\ 
			\bottomrule
	\end{tabular}
	\label{tab:simulation_setup}
\end{table}

\subsubsection{Training algorithm for DS deployment} 
		This algorithm is designed to generate DS enhanced codebooks that can be used for  DS deployment both in the SSLCF framework and the MSLCF framework. The initial DS codebook, which is the conventional RVQ codebook $\mathcal{C}$, is refined through a periodic update mechanism. The overall training and inference process is summarized in \algref{alg:DCT}. The same loss function as in (\ref{loss}) is employed. Additionally, training algorithm for DS deployment incorporates a periodic codebook update mechanism: after a fixed number of epochs, the current model output is used to update the codebook. This updated codebook is evaluated on a validation set, and if it results in a lower validation loss, it replaces the previous version; otherwise, the earlier best is retained. This iterative update guarantees that the DS codebooks consistently improves codewords refinement accuracy.

\section{Simulation Results}\label{Section 5}

\begin{figure}[!t]
	\centering
	\subfigure[mmWave scenario.]{
		\begin{minipage}[b]{ \linewidth}
			\centering
			\includegraphics[width=  0.9\linewidth]{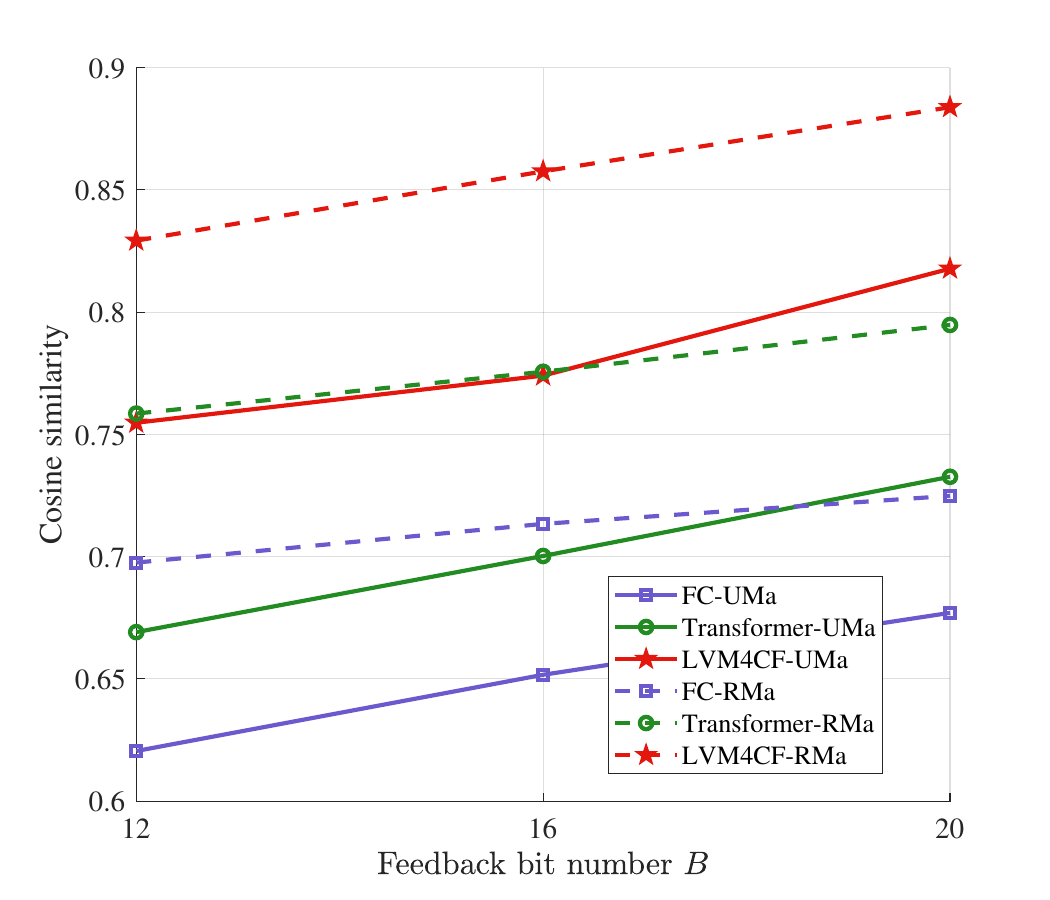}
			\label{fig:mmwave_CosineSimilarity}
		\end{minipage}
	}
    
	\subfigure[sub-6G scenario.]{
		\begin{minipage}[b]{ \linewidth}
			\centering
			\includegraphics[width= 0.9\linewidth]{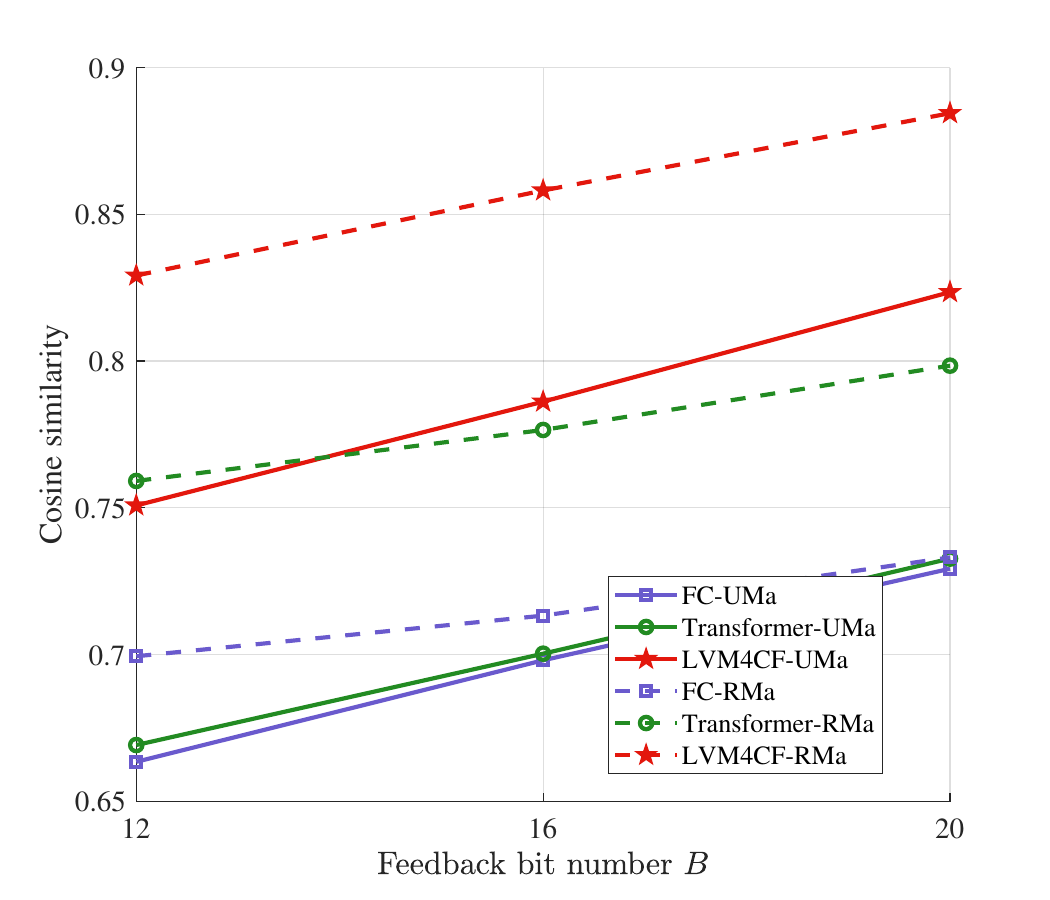}
			\label{fig:sub6G_CosineSimilarity}
		\end{minipage}
	}
	\caption{Cosine similarity vs. the number of feedback bits $B$ (MF-DS).}
	\label{fig:MSLCF_Scenarios}
\end{figure}
    
This section evaluates the performances of the proposed SSLCF and MSLCF frameworks and the LVM4CF network. The parameters used in the simulations are listed in Table \ref{tab:simulation_setup}. 
The channel data are generated using the QuaDRiGa channel simulator \cite{quadriga}, which has been validated through alignment experiments and is officially acknowledged by 3GPP as a reliable tool for channel simulation \cite{3gpp_tr_38_901}. In the SSLCF framework, the dataset comprises $70,000$ training samples, $20,000$ validation samples, and $10,000$ test samples. All samples are synthesized under the same BS and channel environments, ensuring that training, validation, and testing data share a consistent site-specific propagation environment. In contrast, the MSLCF framework utilizes a larger dataset for each large-scale propagation environment (e.g., UMa at $28$ GHz or RMa at $3.5$ GHz), comprising $140,000$ training samples, $40,000$ validation samples, and $20,000$ test samples. These samples are synthesized from $100$ distinct BS deployment scenarios within the same environment, which implies different randomly generated BS locations and surrounding geographical layouts, including variations in scatterer distributions, building placements, and UE positioning. This design enables the MSLCF framework to generalize across a diverse set of site-specific propagation conditions under a common environmental setting, ensuring robustness to deployment variability.

For the LVM4CF backbone, the LVM is based on the open-source code and pretrained weights from \cite{sec3_2}. The feature dimension $D$ is set to 4096, and the first two layers ($L = 2$) of the LVM are deployed for CSI feedback. The simulation setup is designed to reflect realistic channel conditions and system constraints, ensuring that the performance comparisons are both fair and representative of practical deployment scenarios. All networks are trained on 4 NVIDIA 4090 GPUs, and the hyper-parameters are summarized in Table \ref{tab:simulation_setup}.

\begin{figure}[!t]
    \centering
    \includegraphics[width=  0.9\linewidth]{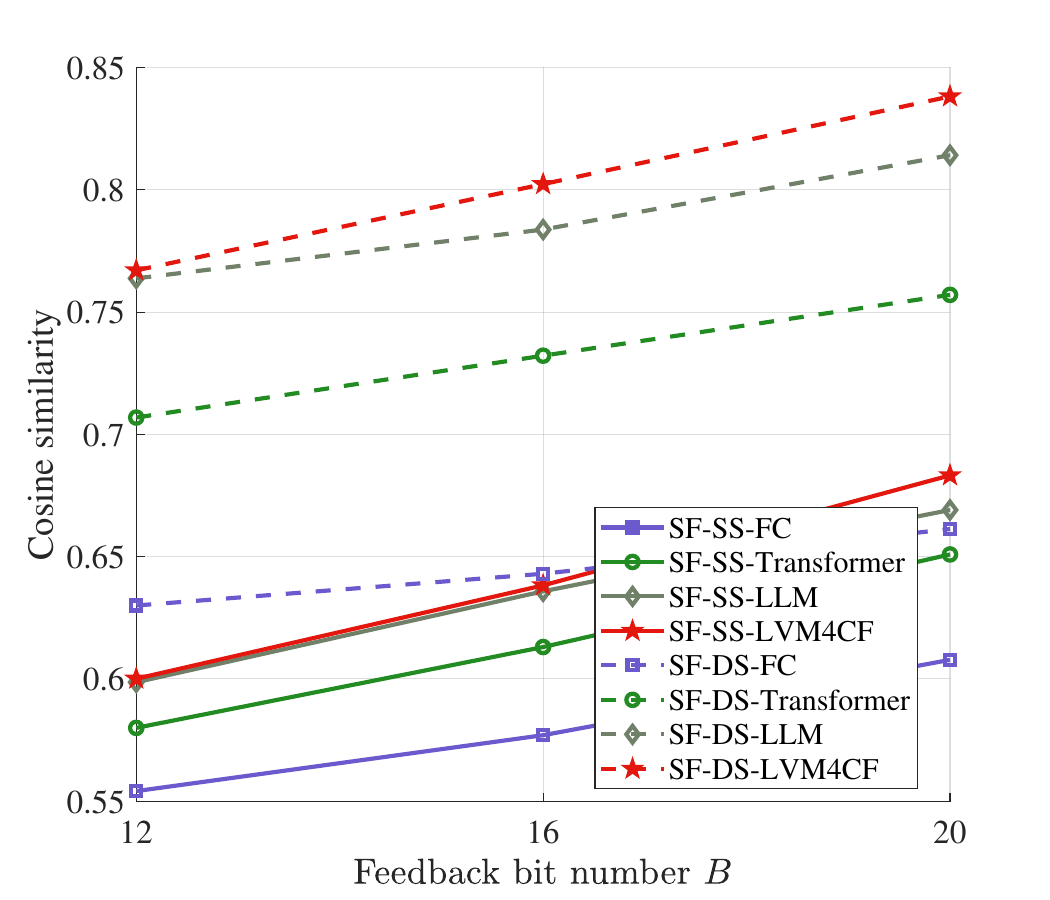}
    \caption{Cosine similarity vs. the number of feedback bits $B$ (SF-SS and SF-DS in UMa-mmWave scenario).}
    \label{fig:BSLCF_BCT_vs_DCT}
\end{figure}

\begin{figure}[!t]
	\centering
	\includegraphics[width=  0.9\linewidth]{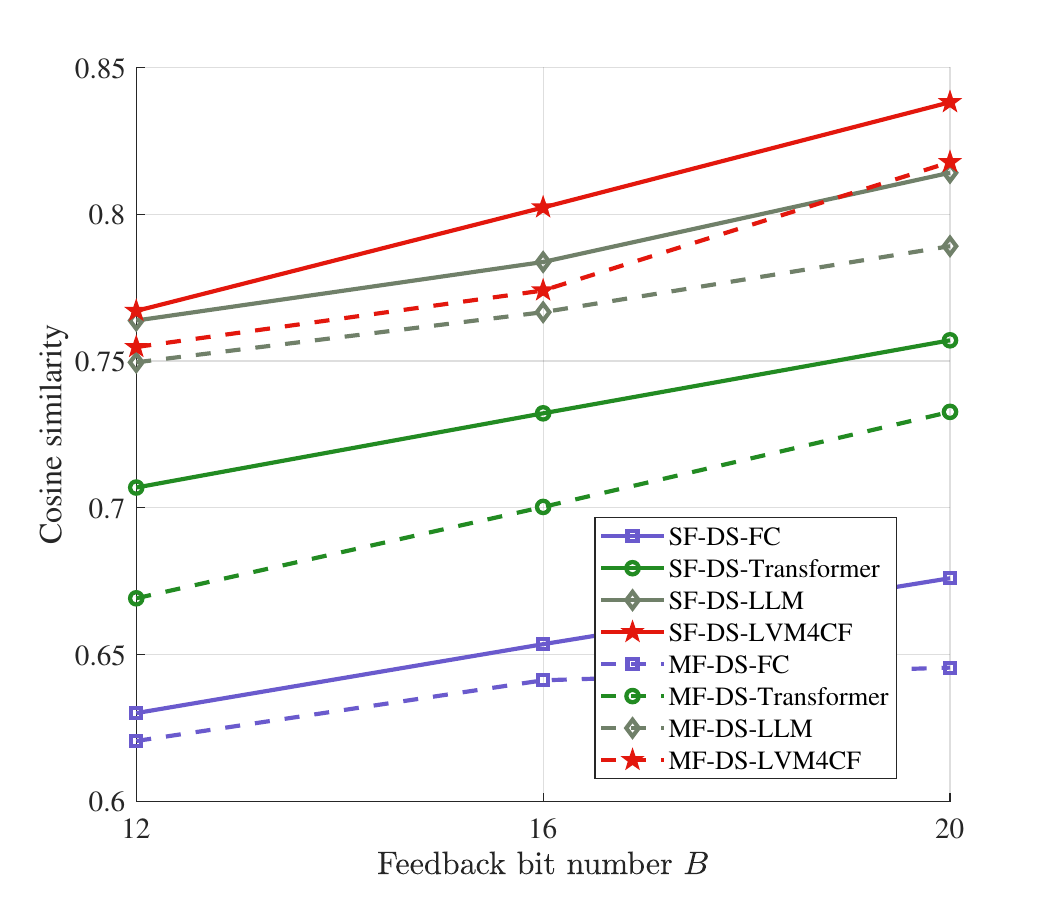}
	\caption{Cosine similarity vs. the number of feedback bits $B$ (SF-DS and MF-DS in UMa-mmWave scenario).}
	\label{fig:BSLCF_vs_MSLCF_UMa28GHz}
\end{figure}

\begin{figure}[!t]
\centering
\subfigure[Network parameters.]{%
\includegraphics[width= 0.9 \linewidth]{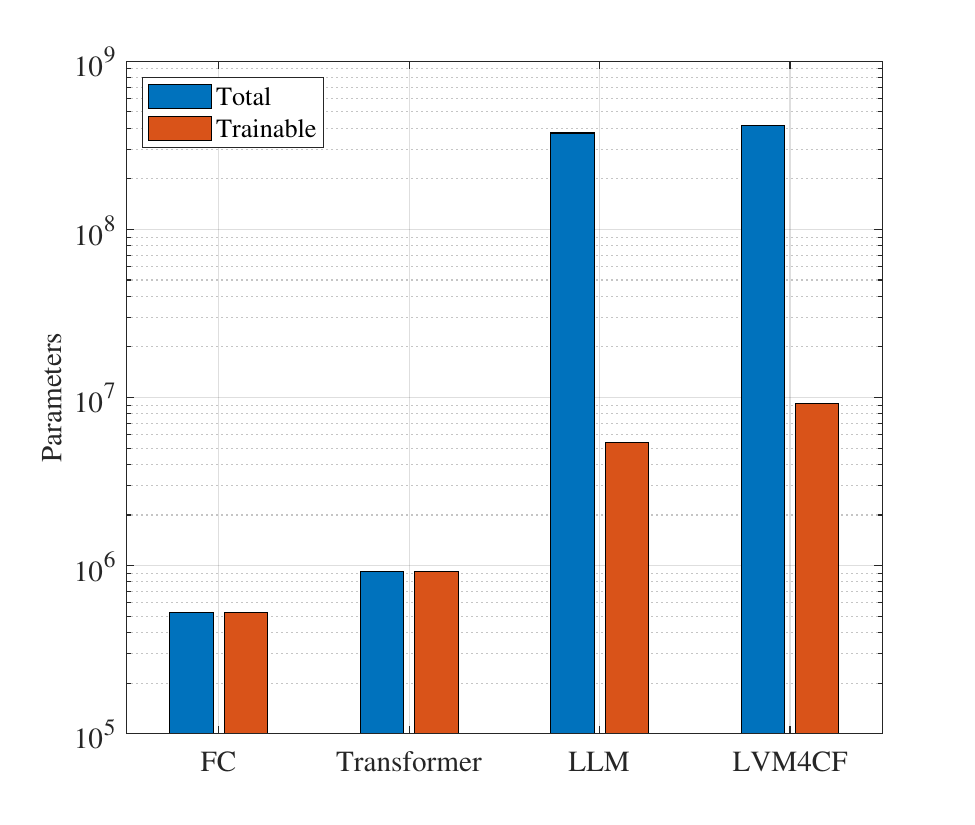}%
\label{fig:network_params}%
}
\subfigure[Running time of CSI feedback.]{%
\includegraphics[width= 0.9 \linewidth]{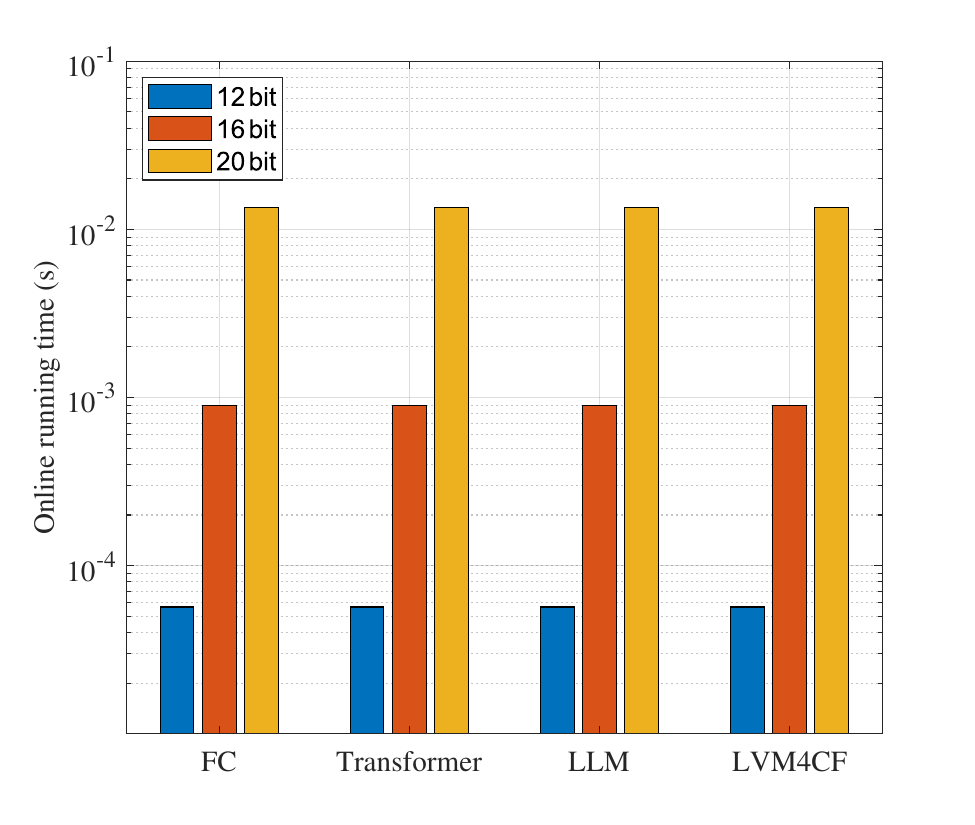}%
\label{fig:feedback_time}%
}
\subfigure[Sum rate, ${\rm SNR}\!=\!20{\rm dB}$.]{%
\includegraphics[width= 0.9 \linewidth]{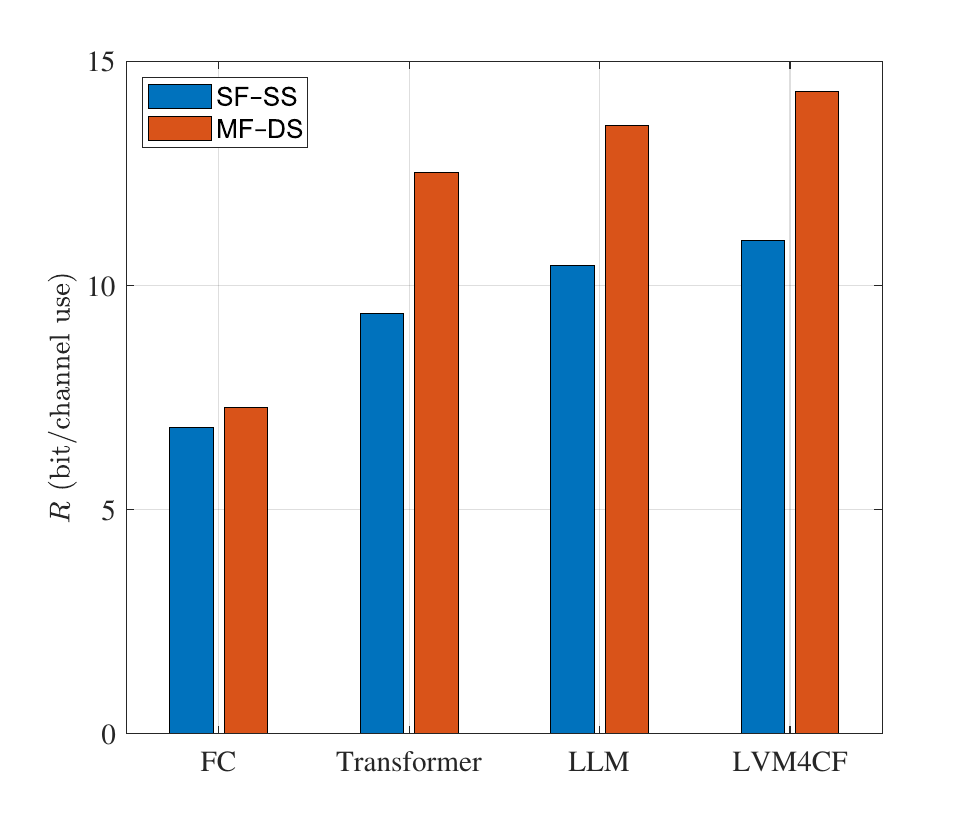}%
\label{fig:average_perf}%
}
\caption{Comparison of model complexity, running time, and communication performance.}
\label{fig:architecture_comparison}
\end{figure}
\begin{figure*}[t]
	\centering
	\subfigure[SS deployment, feedback bit number $B=12$.]{
		\includegraphics[width=0.48\linewidth]{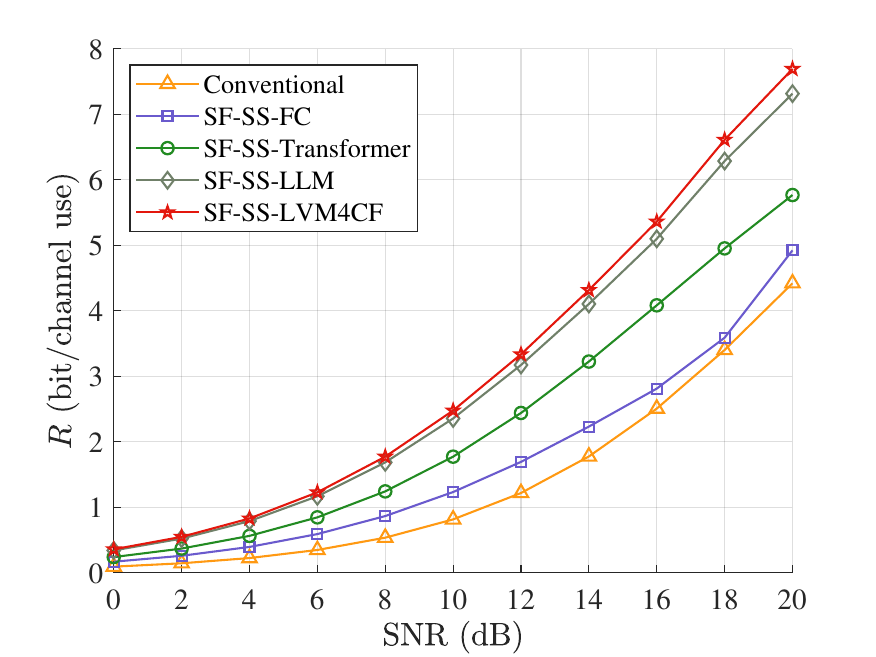}
		\label{subfig:BCT_12bit}
	}
	\subfigure[SS deployment, feedback bit number $B=20$.]{
		\includegraphics[width=0.48\linewidth]{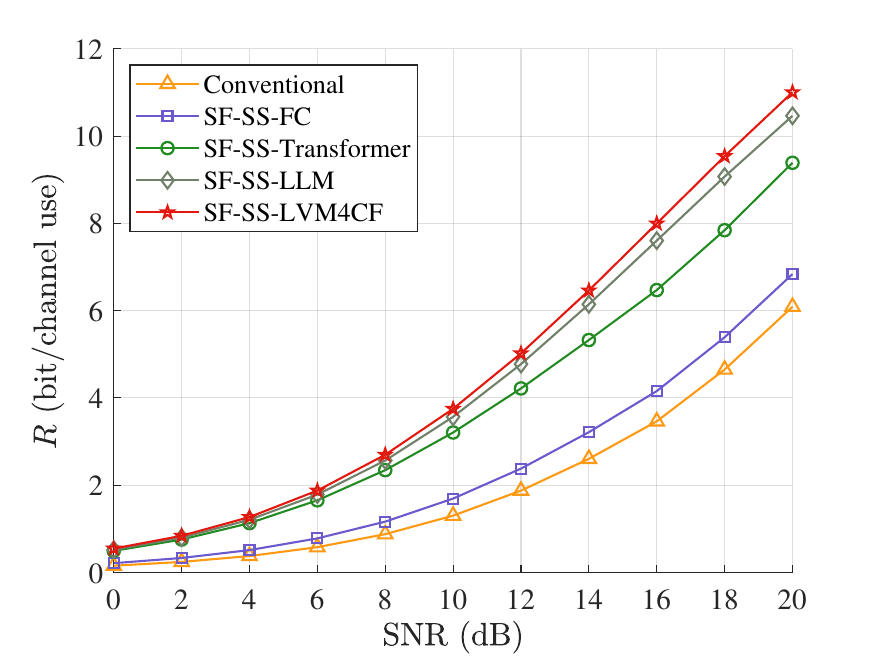}
		\label{subfig:BCT_20bit}
	}
	
	\subfigure[DS deployment, feedback bit number $B=12$.]{
		\includegraphics[width=0.48\linewidth]{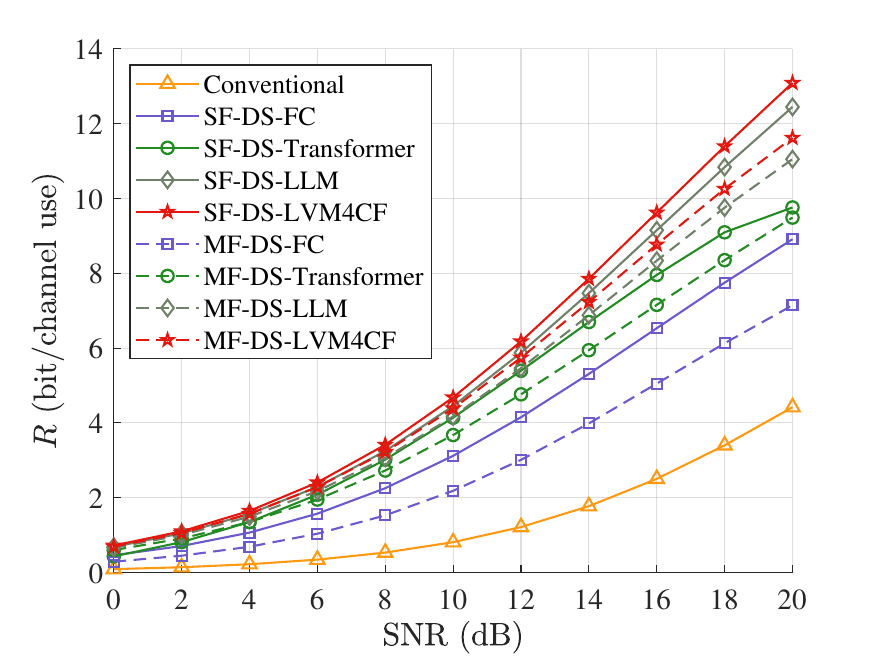}
		\label{subfig:DCT_12bit}
	}
	\subfigure[DS deployment, feedback bit number $B=20$.]{
		\includegraphics[width=0.48\linewidth]{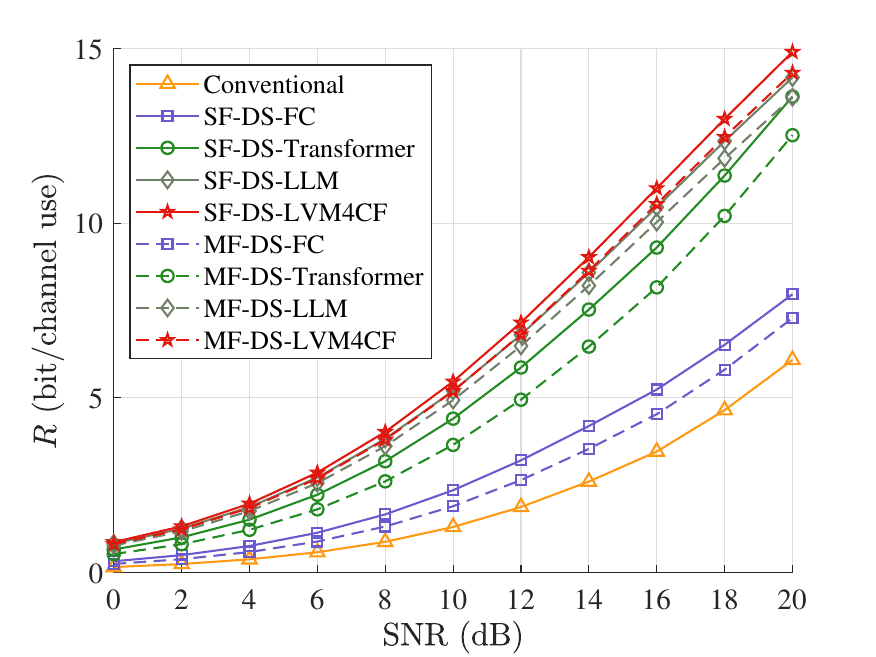}
		\label{subfig:DCT_20bit}
	}
	
	\caption{Sum rate vs. SNR (in UMa-mmWave scenario).}
	\label{fig:sumrate_comparison}
\end{figure*}
The comparisons in this section will cover the following frameworks and deployment strategies:
\begin{itemize}
    \item \textbf{`SF'} and \textbf{`MF'}: The proposed SSLCF framework for site-specific transmissions and the MSLCF framework for generalized multi-scenario transmissions. 

    \item \textbf{`SS'} and \textbf{`DS'}: The SS deployment which deploy the enhanced codebooks for the BS side, and DS deployment which deploy the enhanced codebooks for both the BS and UE sides.
\end{itemize}
On this basis, this section compares the performance of CSI feedback aided by the following network architectures:
\begin{itemize}
\item \textbf{`Conventional'}: Both the UE and the BS use the conventional codebook, i.e., ${\mathcal C}_{\rm u} = {\mathcal C}_{\rm b} = {\mathcal C}$.
\item \textbf{`FC'}: The FC network proposed for codebook-based CSI feedback in \cite{AI4C2F}.
\item \textbf{`Transformer'}: A lightweight Transformer-based model with 2 layers and feature dimension 256 \cite{transformer}.
\item \textbf{`LLM'}: A large language model based on the 12-layer GPT2-XL \cite{gpt}.
\item \textbf{`LVM4CF'}: The proposed LVM4CF network in \secref{network}.
\end{itemize}
Without loss of generality, the RVQ codebook is choosed as the conventional codebook ${\mathcal C}$  \cite{sec2B3_1}.

\figref{fig:BSLCF_BCT_vs_DCT} shows the cosine similarity under SSLCF framework and two different deployment strategies between the predicted codeword and the ground-truth channel. It can be observed that of all methods, LVM4CF achieves the highest similarity, outperforming both traditional and learning-based baselines. Notably, while the Transformer and FC networks improve upon RVQ, their performance is still significantly lower than that of LLM and LVM4CF. This highlights the benefit of using large-scale pre-trained AI models for channel feature learning, especially when domain-specific adaptation is allowed. Additionally, the comparison between SS and DS deployment strategies clearly demonstrates that DS, which jointly optimizes codebooks at both the BS and UE sides, significantly enhances CSI feedback performance. Regarding the MSLCF framework, \figref{fig:MSLCF_Scenarios} illustrates cosine similarity results for mmWave and sub-6G scenarios across urban and rural environments. LVM4CF consistently outperforms all other methods, highlighting its robustness across diverse channel conditions.

\figref{fig:BSLCF_vs_MSLCF_UMa28GHz} compares the CSI feedback performances of the SSLCF and MSLCF frameworks in the UMa-mmWave scenario, evaluated by cosine similarity under varying feedback bit constraints. It can be seen from figure that the proposed LVM4CF achieves superior performance across both frameworks. Particularly, the SSLCF framework consistently yields higher performance than MSLCF, primarily benefiting from its capability to fine-tune the LVM specifically on a site-specific dataset, thereby accurately capturing site-specific channel characteristics. While the MSLCF framework offers broader applicability across multiple BS environments, such generalization inevitably compromises the precision achievable through site-specific adaptation. Nevertheless, despite slightly trailing behind LVM4CF due to the inherent limitations of text-based pre-training data, the LLM still significantly outperforms lightweight models such as FC and Transformer networks. This substantial advantage further highlights the dominance of LAMs within both SSLCF and MSLCF frameworks.
	
\figref{fig:architecture_comparison} compares the number of parameters, CSI feedback runtime, and communication performance, thereby highlighting the advantages and potential of the proposed method. In addition to cosine similarity, we evaluate the impact of CSI feedback accuracy on the overall system performance through sum rate metrics. Without loss of generality, we adopt a multi-user scenario similar to that in \cite{AI4C2F}, i.e., the BS serves 4 UEs using ZF precoding. As shown in \figref{fig:network_params}, both the LLM and LVM exhibit comparable parameter scales, which are significantly larger than those of conventional AI models. However, owing to the offline training and inference mechanism of the proposed framework in \secref{Section 3}, the inherently long inference latency of large AI models does not affect the CSI feedback running time, as demonstrated in \figref{fig:feedback_time}. Furthermore, as illustrated in \figref{fig:average_perf}, the increased parameter count leads to notable improvements in communication performance. These results demonstrate that the proposed framework successfully balances performance and latency, making it a strong candidate for deployment in practical communication systems.


Finally, we provide a more detailed comparison of transmission performance to draw further conclusions.
\figref{fig:sumrate_comparison} compares the sum rate achieved by SF and MF frameworks under both SS and DS deployment strategies with feedback bit constraints of 12 and 20 bits. Specifically, the SS results under 12-bit and 20-bit feedback are shown in \figref{subfig:BCT_12bit} and \figref{subfig:BCT_20bit}, while the DS results are presented in \figref{subfig:DCT_12bit} and \figref{subfig:DCT_20bit}.  It can be seen from the figure that across the entire SNR range, LVM consistently yields the highest sum rate. Although LLM offers competitive performance, it still falls short of LVM, suggesting that vision-based pre-training provides more relevant features for channel representation compared to text-based pre-training. From an SNR perspective, as the SNR increases, interference becomes dominant relative to noise in the sum-rate calculation, making channel accuracy increasingly critical. Consequently, the gap in sum-rate performance among different schemes widens with rising SNR, further underscoring the importance of superior CSI feedback scheme at high SNR.

\section{Conclusion}\label{Section 6}

In this paper, we investigated novel strategies for integrating LAMs into practical CSI feedback mechanisms for FDD massive MIMO systems through offline codebook optimization. Specifically, two frameworks—the SSLCF and the MSLCF—were proposed, leveraging an LVM pre-trained on large-scale image datasets and subsequently fine-tuned on specific CSI datasets. The SSLCF framework generated highly specialized, site-specific enhanced codebooks that substantially improved CSI reconstruction accuracy, while the MSLCF framework created multiple generalized codebooks suitable for diverse propagation environments, balancing accuracy with flexibility and lower signaling overhead. To effectively exploit the structural similarities between CSI and visual data, we introduced LVM4CF, a customized neural network architecture designed explicitly for refining conventional codebook codewords. Two offline training and inference algorithms tailored to SS and DS deployment strategies were also presented, providing flexibility in balancing system performance with practical deployment constraints. Due to the offline training and inference strategy, the increased complexity of large models introduced no additional latency during online feedback, effectively mitigating practical deployment concerns. The results illustrated the practicality and effectiveness of our approach, highlighting a promising pathway for deploying advanced LAMs in future wireless communication systems.

\ifCLASSOPTIONcaptionsoff
\newpage
\fi

\bibliographystyle{IEEEtran}
\bibliography{IEEEfull}

\end{document}